\documentclass[%
 reprint,
 superscriptaddress,
 amsmath,amssymb,
 aps]{revtex4-2}

\usepackage{graphicx}
\usepackage{dcolumn}
\usepackage{bm}
\usepackage[version=4]{mhchem}
\usepackage{physics}
\usepackage{multirow}
\usepackage{siunitx}

\begin{document}

\title{Towards detection of the molecular parity violation in chiral Ru(acac)$_3$ and Os(acac)$_3$}

\author{{Marit R. Fiechter}}
\affiliation{Van Swinderen Institute for Particle Physics and Gravity (VSI), University of Groningen, Groningen, The Netherlands}
\affiliation{ Department of Physics, ETH Z\"urich, Otto-Stern-Weg 1, 8093 Zurich, Switzerland}

\author{Pi A.B. Haase}
\affiliation{Van Swinderen Institute for Particle Physics and Gravity (VSI), University of Groningen, Groningen, The Netherlands}

\author{Nidal Saleh}
\affiliation{Department of Organic Chemistry, University of Geneva, Quai Ernest Ansermet 30, 1211 Geneva 4, Switzerland}
\affiliation{Univ Rennes, CNRS, ISCR-UMR 6226, Campus de Beaulieu, 35042 Rennes Cedex, France}

\author{Pascale Soulard}
\affiliation{Sorbonne Université, CNRS, UMR 8233, MONARIS, Case courrier 49, 4 place Jussieu, F-75005, Paris, France
}

\author{Beno\^it Tremblay}
\affiliation{Sorbonne Université, CNRS, UMR 8233, MONARIS, Case courrier 49, 4 place Jussieu, F-75005, Paris, France
}

\author{Remco W.A. Havenith}
\affiliation{Zernike Institute for Advanced Materials, University of Groningen, Nijenborgh 4, 9747 AG, Groningen, The Netherlands}
\affiliation{Stratingh Institute for Chemistry, University of Groningen, Nijenborgh 4, 9747 AG, Groningen, The Netherlands}
\affiliation{Ghent Quantum Chemistry Group, Department of Inorganic and Physical Chemistry, Ghent University, Krijgslaan 281 (S3), B‐9000 Ghent, Belgium}

\author{Rob G.E. Timmermans}
\affiliation{Van Swinderen Institute for Particle Physics and Gravity (VSI), University of Groningen, Groningen, The Netherlands}

\author{Peter Schwerdtfeger}
\affiliation{Centre for Theoretical Chemistry and Physics, The New Zealand Institute for Advanced Study, Massey University, 0745 Auckland, New Zealand}

\author{Jeanne Crassous}
\affiliation{Univ Rennes, CNRS, ISCR-UMR 6226, Campus de Beaulieu, 35042 Rennes Cedex, France}

\author{Beno\^it Darqui\'e}
\affiliation{Laboratoire de Physique des Lasers, Université Sorbonne Paris Nord, CNRS, Villetaneuse, France}

\author{Luk\'a\v s F. Pa\v steka}
\affiliation{Department of Physical and Theoretical Chemistry, Faculty of Natural Sciences, Comenius University, Ilkovičova 6, 84215 Bratislava, Slovakia}

\author{Anastasia Borschevsky}
\email{a.borschevsky@rug.nl}
\affiliation{Van Swinderen Institute for Particle Physics and Gravity (VSI), University of Groningen, Groningen, The Netherlands}

\date{\today}

\begin{abstract} 

We present a theory-experiment investigation of the helically chiral compounds Ru(acac)$_3$ and Os(acac)$_3$ as candidates for the next-generation experiments for detection of molecular parity violation (PV) in vibrational spectra. We used state-of-the-art relativistic calculations to identify optimal vibrational modes with expected PV effects exceeding by up to two orders of magnitude the projected instrumental sensitivity of the experiment under construction at the Laboratoire de Physique des Lasers in Paris. High-resolution measurements of the vibrational spectrum of Ru(acac)$_3$ carried out as the first steps towards the planned experiment are presented.

\end{abstract}

\maketitle

\emph{Introduction. --}
Parity violation (PV) through the weak interaction was
first predicted in 1956 \cite{leeyang}, immediately after observed in nuclear physics \cite{wu1957experimental,PhysRev.105.1415}, 
and later in atomic physics \cite{barko1978measurement,bouchiat1982observation,gilbert1985measurement,drell1984parity}.
In chiral molecules, it is predicted to result in a tiny energy difference between enantiomers which, if large enough, may provide the bias needed to seed the observed biomolecular homochirality, \emph{i.e.} the fact that chiral molecules usually occur in only one enantiomeric form in nature  \cite{yamagata1966hypothesis,mason1984origins,tranter1985parity}. This PV energy difference can also serve as a sensitive probe of the electroweak interactions (to naturally complement low-energy tests using atoms) and of new physics beyond the standard model. It is predicted to be particularly sensitive to parity-violating cosmic fields, which are invoked in different models for cold dark matter or in the Lorentz-invariance violating standard model extensions \cite{gaul2020chiral}.

Over the past decades, various experiments have been proposed to observe parity violation in chiral molecules, including measurements of PV frequency shifts in NMR spectroscopy (\emph{e.g.} \cite{barra1986parity,robert2001nmr,bast2006parity,eills2017measuring}), measurements of the time-dependence of optical activity \cite{harris1978quantum}, and direct measurement of the absolute PV energy shift of the electronic ground state \cite{quack1986measurement}; see 
various reviews in this field \cite{berger2019parity,pvinbookberger,Schwerdtfeger2010}.
However, for none of the aforementioned experimental schemes, tight experimental upper bounds have been reported yet \cite{berger2019parity}; this has only been accomplished in the measurements of the PV shift of vibrational frequencies, performed at the Laboratoire de Physique des Lasers (LPL) in Paris \cite{daussy1999limit,ziskind2002improved,darquie2010progress,tokunaga2013probing,cournol2019new}, using ultra-precise mid-infrared molecular spectroscopy experiments.   

Several molecules have been considered as candidates for experiments at LPL. The first experiments were performed on the C--F stretch vibration in CHFClBr lying conveniently in the CO$_2$ laser frequency range \cite{kompanets1976narrow,daussy1999limit,ziskind2002improved,Schwerdtfeger2002}, but led to a non-detection, with an upper limit of $\Delta \nu^\text{PV}/\nu = 2.5 \times 10^{-13}$, with $\nu$ the vibrational transition frequency and $\Delta\nu^\text{PV}$ the parity violating frequency difference between enantiomers. Later, \textit{ab initio} calculations predicted the PV shift for this transition to be 3 to 4 orders of magnitude smaller \cite{quack2003combined,schwerdtfeger2005relativistic,thierfelder2010relativistic,RauSch2021}. Another fluorohalomethane that has been under investigation, CHFClI, is predicted to have a larger PV shift, but is not stable enough for measurements \cite{darquie2010progress,soulard2006chlorofluoroiodomethane}. 
Other candidate molecules include \ce{SeOClI} \cite{figgen2008seocli}, \ce{N$\equiv$WHClI} \cite{figgen2010nwhcli}, and \ce{N$\equiv$UHXY} \cite{wormit2014strong} (X, Y = F, Cl, Br, I). These systems were predicted to possess vibrational transitions with progressively larger PV shifts, as large as several tens of Hz (or $\Delta \nu^\text{PV}/\nu \sim 10^{-13}$) in \ce{N$\equiv$UHFI}. So far, the synthesis of this type of compounds has not been reported. More recently, attention has turned to chiral oxorhenium complexes \cite{stoeffler2011high,saleh2013chiral,saleh2018oxorhenium}, but bringing these into the gas phase, which is required for high-precision spectroscopy, remains a challenge given their low stability.

\begin{figure}
	\centering
	\includegraphics[width=2.5in]{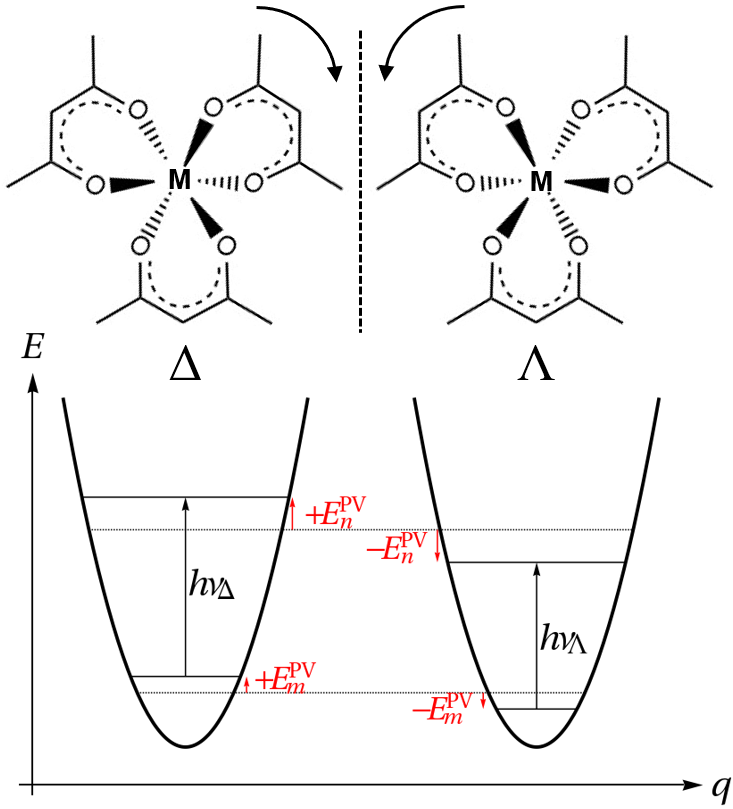}
	\caption{Chemical structures of the $\Delta$- and $\Lambda$-M(acac)$_3$ (M~=~Ru, Os) with the corresponding transition frequencies $\nu_\Delta$ and $\nu_\Lambda$.}
	\label{fig:Ruacac} 
\end{figure}

In this paper, we investigate a substantially distinct class of highly promising molecules, namely chiral M(acac)$_3$ complexes (where M is a metal and acac stands for acetylacetonate, see Figure~\ref{fig:Ruacac}). Specifically, we focus on Ru(acac)$_3$ and its heavier homologue Os(acac)$_3$. Ruthenium ($Z = 44$) and osmium ($Z = 76$) have reasonably heavy nuclei, so that in accord with the proposed  $Z^5$-scaling law \cite{zeldovich1977jetp,bast2011analysis,wormit2014strong} we expect these systems to experience a large absolute PV energy shift. Compared to previously proposed species, these molecules exhibit widely different chemical properties – in particular their propeller-like chiral topology (see Figure~\ref{fig:Ruacac}) – and physico-chemical properties allowing a number of the above-mentioned limitations to be overcome. In support of M(acac)$_3$ molecules, we note that compared to the more exotic species presented above: 
(i) these are well-known classical archetype systems in organometallic chemistry \cite{sato_effects_2007},
(ii) they feature a relatively high volatility,
(iii) with Ru(acac)$_3$ being commercially available in its racemic form and being resolvable into pure $\Delta$ and $\Lambda$ enantiomers at gram scales \cite{drake_optical_1983}. 
In fact, we have recently demonstrated that Ru(acac)$_3$ is stable and robust under evaporation by heating up to at least $200^{\circ}$C, and that it can easily be brought into the gas phase \cite{darquie_pecd_2020}. This allowed us to seed it in a molecular beam (see \cite{darquie_pecd_2020}) and in a solid matrix of neon at 3K, allowing preliminary mid-infrared Fourier transform spectroscopic investigations, as reported in the present work. We thus demonstrate an outstanding level of control of a heavy organo-metallic chiral candidate species for measuring PV. This study leads us to expect that this molecule will survive laser ablation and allow the production of cold and slow gas samples via buffer-gas cooling in a cryogenic chamber, a method that we have recently demonstrated with other organo-metallic species \cite{cournol2019new,asselin_characterising_2017,Tokunaga2017}. Buffer-gas-cooled beams, the latest molecular beam technology at the heart of the LPL apparatus \cite{cournol2019new}, will provide the low temperature, low speed, and high intensity needed for measuring PV. All in all, these factors make Ru(acac)$_3$ a highly attractive candidate for precision gas-phase spectroscopy experiments.

The heavier homologue, Os(acac)$_3$, is not commercially available but can be synthesised \cite{dallmann1998darstellung}. Its resolution into pure enantiomers is to be investigated in future experiments.
Its higher atomic number will lead to larger vibrational PV shifts \cite{wormit2014strong}, making it a promising alternative to Ru(acac)$_3$.

In this work, we perform theoretical investigations of the magnitude of the PV shifts in Ru(acac)$_3$ and Os(acac)$_3$. Precise spectroscopic data tends to be scarce or simply non-existent for such species. Here we perform first high-resolution spectroscopic investigations of Ru(acac)$_3$ that in combination with the calculations allow us to select the most appropriate transitions for measuring parity violation. Such \emph{a priori} investigations are l
also crucial to determine the frequency range for the metrology-grade laser system to be built and tuned at LPL, as the size of the PV vibrational shift can vary by over an order of magnitude, depending on the vibrational mode \cite{RauSch2021}.

\emph{Methodology. --}
According to the standard model of particle physics, the dominant $P$-odd contribution of the weak force to the molecular Hamiltonian is due to vector-nucleonic--axial-vector-electronic coupling. In the low-energy limit, the following nuclear spin-independent effective Hamiltonian can be derived from the standard model Lagrangian (see \emph{e.g.} Refs. \cite{pvinbookberger,nahrwold2011electroweak}):
\begin{equation} \label{eq:HamiltonianPV}
\hat{H}^\text{PV}= \frac{G_\text{F}}{2\sqrt{2}} \sum_j^\text{electrons}\sum_A^\text{nuclei}
Q_\text{W}(A)\gamma_j^5 \rho(\mathbf{r}_j-\mathbf{r}_A),
\end{equation} 
which is compatible with the usual four-component framework for relativistic quantum chemical calculations. In this equation, $G_\text{F}= 2.22255\times 10^{-14}$
a.u. 
stands for the Fermi coupling constant; the weak charge of nucleus $A$ is given by $Q_\text{W}(A)=[(1-4\sin^2\theta_\text{W})Z-N]$, where $\theta_\text{W}$ is the weak mixing angle, and $Z$ and $N$ are the number of protons and neutrons, respectively; $\rho(\mathbf{r})$ stands for the nuclear density; and the fifth gamma matrix can be written 
in terms of the Dirac matrices
as $\gamma^5=-i\gamma^0\gamma^1\gamma^2\gamma^3$.
The Hamiltonian 
will yield a contribution to the energy that is positive for one enantiomer, and negative for the other.

In this work, we calculate the PV difference in the vibrational transition frequencies between the two enantiomers as illustrated in Figure~\ref{fig:Ruacac}.
The computational procedure is structured as follows. 
The molecular geometries are optimized at the density functional theory (DFT) level, using an effective core potential on the central metal atom to account for scalar relativistic effects. The normal modes and corresponding frequencies are calculated in a harmonic frequency analysis.
Subsequently, several normal modes are selected for further fully relativistic calculations; details of the selection criteria are presented later in the text. 
For each chosen mode, relativistic DFT calculations are performed 
along the normal mode to obtain the PV potential $V^\text{PV}(q)$ as a function of the normal coordinate, $q$.
Full details of the computational setup used in this work (programs, DFT functionals, basis sets etc.) as well as the investigation of the robustness of this particular computational configuration
can be found in the Supplementary Material \cite{SI}.

Next, the vibrational wavefunctions $\ket{n}$ are determined by solving the vibrational Schr\"{o}dinger equation numerically following the Numerov--Cooley procedure, for the potential obtained along the normal mode \cite{noumerov1924method,cooley1961improved,radovancode}. Then, the shift of a vibrational level $n$ can be calculated as 
\begin{equation} \label{eq:E_pv_int}
  E^\text{PV}_n=\bra{n}V^\text{PV}(q)\ket{n}.
\end{equation}
From this we find the PV frequency difference for a transition from level $m$ to level $n$ between the left- and right-handed form of the molecule:
\begin{equation}
    \Delta \nu^\text{PV}_{m\rightarrow n}=\frac{2}{h}(E^\text{PV}_n-E^\text{PV}_m),
\end{equation}
with $h$ being Planck's constant, and where the factor of two arises from the fact that the energy of one enantiomer is shifted up by the PV effects, while that of the other is shifted down by the same amount.

In this work, we define the sign of this frequency difference via
\begin{equation}
    \Delta \nu^\text{PV}_{m\rightarrow n} = \nu_\Delta - \nu_\Lambda = \nu_\Delta^\text{PV} - \nu_\Lambda^\text{PV} = 2 \nu_\Delta^\text{PV},
\end{equation}
where $\nu_\Delta$ and $ \nu_\Lambda$ are the $m \rightarrow n$ transition frequencies and $\nu_\Delta^\text{PV}$ and $ \nu_\Lambda^\text{PV}=-\nu_\Delta^\text{PV}$ stand for the frequency shifts of these transitions in the $\Delta$ and $\Lambda$ enantiomer, respectively (see Figure~\ref{fig:Ruacac}).

As the relativistic calculations are rather expensive computationally, instead of calculating the PV shifts for all normal modes, we select a specific subset of modes for our investigation 
based on the outcome of the vibrational analysis. In order to have a large differential PV shift, the parity violating energy should vary significantly over the range of a vibration \cite{wormit2014strong}. 
The PV effects scale steeply with $Z$ and thus the metal atom will contribute the most to the total PV energy difference. In order to achieve a large change in the electronic density in the vicinity of the metal atom over the course of a vibration, we look for modes with a large change in the position of the surrounding oxygen atoms.
As a measure hereof, we take the sum of the moduli of the M--O displacements along the normal mode $q$:
\begin{equation}\label{MOdisp}
    \sum_{i=1}^6\sqrt{(\Delta x_{\text{M--O},i})^2+(\Delta y_{\text{M--O},i})^2+(\Delta z_{\text{M--O},i})^2}
\end{equation}
where 
$(\Delta x_{\text{M--O},i}$, $\Delta y_{\text{M--O},i}$, $\Delta z_{\text{M--O},i})^\text{T} = \vec{d}_{M}-\vec{d}_{O,i}$, \emph{i.e.} the differences of the metal and $i$-th oxygen displacement coordinates associated with normal mode $q$ (with normalised corresponding displacement vectors).

\emph{Results: \ce{Ru(acac)_3}. -- }
To carry out the first high-resolution vibrational spectroscopic investigation of \ce{Ru(acac)3}, we synthesized grams of pure $\Lambda$ and $\Delta$ enantiomers (following the recipe detailed in the Supplementary Material \cite{SI}) and recorded the Fourier transform infrared spectrum of $\Lambda$-\ce{Ru(acac)3} trapped in solid neon at 3~K. Such low-temperature matrix-isolation measurements are not muddled by rotations, which are mostly inhibited, and exhibit narrower bands than the more traditional room-temperature studies in the liquid or solid phase. Importantly, the obtained vibrational band centres are typically shifted by only a few wavenumbers (0.3\% of the transition frequency at most \cite{SI}) with respect to the gas-phase conditions required for the PV measurements, a level of uncertainty that the current theory cannot provide. The quantum cascade lasers (QCLs) that will be used in the PV measurements typically cover a few wavenumbers. Thus, in combination with theoretical guidance on the optimal vibrational modes for measurements, this precursor spectroscopic characterisation is crucial for designing the laser system.

\begin{figure}
	\centering
	\includegraphics[width=0.49\textwidth]{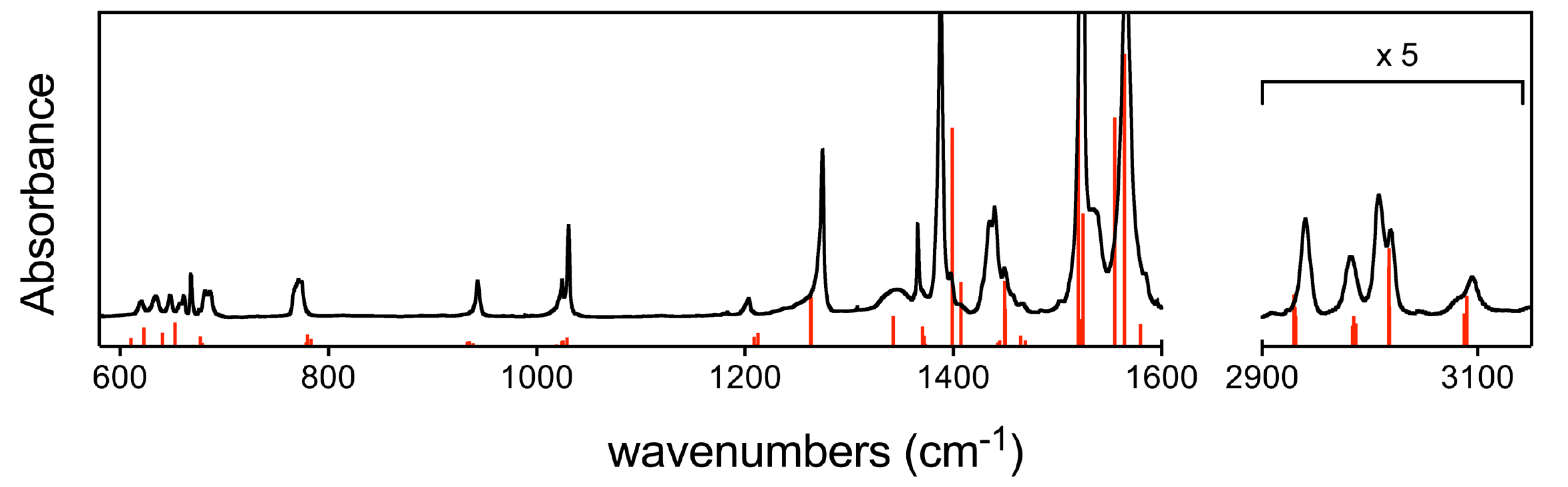}
	\caption{Comparison between the calculated vibrational spectrum (red lines) of \ce{Ru(acac)3} from the geometry optimisation and the infrared Fourier transform spectrum recorded in solid neon at 3~K (above 2800 cm$^{-1}$ intensities and absorbance were multiplied by a factor of 5). The calculated harmonic frequencies greater (resp. smaller) than 1700 cm$^{-1}$ were scaled with a scaling factor of 0.964 (resp. 0.98) for comparison with the measured spectrum. The baseline of the experimental spectrum is vertically shifted for clarity. }
	\label{fig:Ru(acac)3spectrum} 
\end{figure}

Figure~\ref{fig:Ru(acac)3spectrum} compares the calculated harmonic frequencies with the matrix-isolation measurement. Details on the spectroscopy, and in particular on the assignment of the observed bands to the corresponding internal modes, can be found in the Supplementary Material \cite{SI}. Overall, the calculated and the measured spectra are in very good agreement.

\begin{figure}
	\centering
	\includegraphics[width=0.48\textwidth]{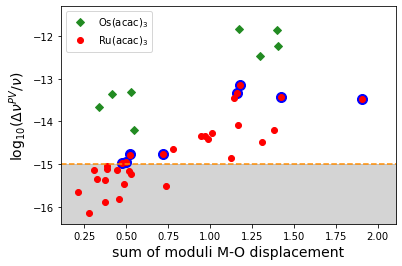}
	\caption{Calculated relative PV frequency shifts ($\Delta\nu^\text{PV}/\nu$) of several vibrational transitions in Ru(acac)$_3$ and Os(acac)$_3$ as a function of the indicator (sum of moduli of Ru--O or Os--O displacements, see text). The red dots represent Ru(acac)$_3$, with the larger dots highlighted in blue corresponding to the selected normal modes shown in Table~\ref{tab:PVresults}; the green diamonds represent the corresponding normal modes of Os(acac)$_3$ shown in Table~\ref{tab:PVresultsOs}. The orange dotted horizontal line is the expected sensitivity attainable by ultra-high resolution vibrational spectroscopy, bordering the grey zone inaccessible to measurements.}
	\label{fig:rmsRu} 
\end{figure}

The calculated PV shifts are plotted against the indicator of the M--O displacement \eqref{MOdisp} in Figure~\ref{fig:rmsRu} for Ru(acac)$_3$. A clear correlation between the magnitude of the displacement and the size of the calculated PV shift can be observed. This correlation will guide the search for viable normal modes in similar compounds.

In addition to the magnitude of the PV frequency shifts, a number of experimental considerations have to be taken into account when selecting the vibrational modes for measurements. The intensity of the selected mode should be high enough to make the measurement feasible and its frequency should lie in a range accessible to current laser technologies. The group at LPL in Paris has \ce{CO2} laters \cite{chanteau_mid-infrared_2013} and QCLs  \cite{cournol2019new,sow_widely_2014,argence_quantum_2015,santagata_high-precision_2019} of record frequency stabilities and accuracies necessary for measuring the tiniest PV frequency differences. The \ce{CO2} lasers, until recently the only available stable sources for precise mid-infrared spectroscopy, span the 9--11~$\mu$m range. Mid-infrared QCLs are commercially available in the 4--13~$\mu$m range and more sporadically up to 17~$\mu$m. 

Based on the above criteria (in terms of a large predicted PV shift, intensity, and desired wavelength range) 
we display a number of promising normal modes in Table~\ref{tab:PVresults}. All these modes have a relative PV frequency shift $\Delta \nu^\text{PV}/\nu$ of $10^{-15}$ or above, which is the sensitivity aimed for in the LPL experiment. For comparison, the PV shift of CHFClBr, a molecule on which much of the experimental work has been conducted so far, has a predicted $\Delta \nu^\text{PV}/\nu \approx -8\times10^{-17}$  \cite{quack2003combined,schwerdtfeger2005relativistic}. 

Particularly interesting for the experiment are thus normal modes 52 and 53 as they are in  the laser window currently available at LPL, are reasonably infrared active, and have a predicted PV shift on the 10$^{-15}$ level. The C--O stretch vibrational modes 100 and 102 look even more promising, given their remarkably high intensity, their twice higher predicted relative PV shift, and the commercial availability of QCLs in the corresponding spectral window. Finally, normal modes 17, 19, 20 and 29 should not be overlooked, not only because of their record 10$^{-14}$ relative predicted shift, but also because of their lower frequencies, which may prevent the onset of intramolecular vibrational energy redistribution that could obscure the spectra at higher frequencies. However, proper radiation sources are still unavailable in this spectral window.

\setlength{\tabcolsep}{8pt}
\begin{table}
\small
\begin{tabular}{ccccc}
\hline 
normal & $\nu$ & IR int. & $\Delta\nu^\text{PV}$ & \multirow{2}{*}{$\displaystyle\left|\frac{\Delta\nu^\text{PV}}{\nu}\right|$}   \\
mode&[cm$^{-1}$]&[km/mol]&[mHz]&\\
\hline 
\hphantom{1}17   & \hphantom{1}182    & 0.009 & --449  & 7.2$\times 10^{-14}$ \\
\hphantom{1}19   & \hphantom{1}201    & 1.718 & --298  & 4.6$\times 10^{-14}$ \\
\hphantom{1}20   & \hphantom{1}223    & 0.065 & \hphantom{--}279   & 3.8$\times 10^{-14}$  \\
\hphantom{1}29   & \hphantom{1}327    & 7.884 & \hphantom{--}325   & 3.4$\times 10^{-14}$  \\
\hphantom{1}52   & \hphantom{1}953    & 9.564 & \hphantom{1}--30   & 1.0$\times 10^{-15}$ \\
\hphantom{1}53   & \hphantom{1}954    & 1.793 & \hphantom{1}--33   & 1.1$\times 10^{-15}$ \\
100  &            1586    & 453.5 & \hphantom{1}--83   & 1.7$\times 10^{-15}$  \\
102  &            1612    & 44.22 & --111  & 2.3$\times 10^{-15}$ \\ 
\hline 
\end{tabular}
\caption{Selected normal modes of Ru(acac)$_3$ promising for PV measurements: possessing either particularly large predicted PV shifts on the $10^{-14}$ level (modes 17-29), or a large shift in combination with a frequency in the range of current laser systems (modes 52 and 53), or very large infrared intensity (IR int.) (modes  100 and 102). The harmonic vibrational frequencies ($\nu$) were obtained from the frequency analysis. }
\label{tab:PVresults} 
\end{table}

\emph{Results: \ce{Os(acac)3}. -- }
Our calculated frequencies for Os(acac)$_3$ are in good agreement with the vibrational spectrum recorded by Dallmann and Preetz \cite{dallmann1998darstellung}. Details on this comparison can be found in the Supplementary Material \cite{SI}.

\begin{table*}[t]
\begin{tabular}{cccccccc}
\hline
\multirow{2}{*}{mode(Os)} &  $\nu$ & {IR int.} & $\Delta\nu^\text{PV}$  & \multirow{2}{*}{$\displaystyle\left|\frac{\Delta\nu^\text{PV}}{\nu}\right|$} & \multirow{2}{*}{mode(Ru)} & \multirow{2}{*}{$\phi$} & \multirow{2}{*}{$\displaystyle\frac{\Delta\nu^\text{PV}\text{(Os)}}{\Delta\nu^\text{PV}\text{(Ru)}}$}   \\
&[cm$^{-1}$]&[km/mol]&[Hz]&&&&\\
\hline
\hphantom{1}16   & \hphantom{1}191 & 0.091 & --9.72   & 1.5$\times 10^{-12}$& \hphantom{1}17    & 0.996 & 21.7\\
\hphantom{1}19  & \hphantom{1}211 & 2.356 & --9.59  &  1.4$\times 10^{-12}$ & \hphantom{1}19    & 0.961 & 32.2 \\
\hphantom{1}20   & \hphantom{1}224 & 0.013 & \hphantom{--}4.30     & 5.9$\times 10^{-13}$ & \hphantom{1}20    & 0.948& 15.4\\
\hphantom{1}29  & \hphantom{1}308 & 2.392 & \hphantom{--}3.09     & 3.4$\times 10^{-13}$ & \hphantom{1}29    & 0.838 & \hphantom{1}9.5\\
\hphantom{1}52   & \hphantom{1}952 & 0.464 & --1.47  & 5.0$\times 10^{-14}$   & \hphantom{1}52    & 0.830 & 48.4 \\
\hphantom{1}53   & \hphantom{1}954 & 1.272 & --1.32   & 4.5$\times 10^{-14}$ & \hphantom{1}53    & 0.831& 40.1\\
            100 &          1563    & 245.6 & --1.04   & 2.2$\times 10^{-14}$ &  100     & 0.955& 12.7\\
            102 &          1589    & 102.2 & --0.31   & 6.3$\times 10^{-15}$ & 102     & 0.985 & \hphantom{1}3.6 \\
\hline
\end{tabular}
\caption{PV shifts of vibrational normal modes in \ce{Os(acac)3} and a comparison to \ce{Ru(acac)3}. Normal modes in the two compounds were matched to each other according to their overlap $\phi$ as defined in Eq. 
(1) in the Supplementary Material \cite{SI}.
The harmonic vibrational frequencies ($\nu$) were obtained from the frequency analysis.}
\label{tab:PVresultsOs} 
\end{table*}

For \ce{Os(acac)3}, calculations were performed for the normal modes that correspond to those of \ce{Ru(acac)3} displayed in Table~\ref{tab:PVresults} (details in the Supplementary Material \cite{SI}). 
%
The results 
for Os(acac)$_3$ are presented in Table~\ref{tab:PVresultsOs} and in Figure~\ref{fig:rmsRu}. For all of the presented normal modes, the signs of the PV shifts are the same for the Ru and Os complexes. This emphasizes the similarity of the vibrational modes between the two complexes and the robustness of the PV shift under slight geometric changes.
A striking observation is that the relative PV shifts ${|\Delta \nu^\text{PV}/\nu|} $  in \ce{Os(acac)3} are more than an order of magnitude larger than those in \ce{Ru(acac)3}, reaching $10^{-12}$ levels. A closer look at the ratio of absolute shifts 
reveals, for some of the transitions, a significant enhancement beyond the $Z^5$ scaling, which would amount to $\Delta\nu^\text{PV}_\text{\ce{Os(acac)3}}/\Delta\nu^\text{PV}_\text{\ce{Ru(acac)3}}\approx 15.4$. This is not entirely unexpected, as the $Z^5$ dependence was derived for the absolute PV energy shifts \cite{zeldovich1977jetp} rather than for vibrational transitions; furthermore, similar beyond-the-$Z^5$ scaling was observed in chiral uranium compounds by Wormit \textit{et al.} \cite{wormit2014strong}. 

Here, it has very favourable consequences for the experiment; the two transitions in Table~\ref{tab:PVresultsOs} that lie in the laser window currently available at LPL (modes 52 and 53) experience an enhancement of a factor of 48 and 40, respectively, when changing from \ce{Ru(acac)3} to \ce{Os(acac)3}, significantly larger than the $Z^5$ scaling used for a rough estimate would predict. This enhancement pushes the $|\Delta \nu^\text{PV}/\nu|$ of these modes into the  $\sim 10^{-14}$ regime, making it much easier to detect at the level of sensitivity already demonstrated at LPL \cite{darquie2010progress,shelkovnikov_stability_2008}. This finding provides us with a strong motivation to synthesize this compound and bring it into the gas phase.

\emph{Conclusions. -- } We have calculated the PV vibrational frequency shifts for a selection of normal modes in Ru(acac)$_3$, a highly promising candidate species for the first detection of parity violation in molecules. 
We have derived a simple scheme for identifying the most promising vibrational modes that allowed us to pinpoint of transitions with exceptionally large PV shifts of hundreds  of mHz (corresponding to relative frequency shifts on the order of $10^{-15}-10^{-14}$). 
These effects are well within the projected sensitivity of the experiment being built at LPL. 
From the point of view of experimental control, Ru(acac)$_3$ benefits from many advantages compared to previously proposed species. It is a robust, readily available archetypal molecule in organic chemistry, and the first heavy chiral candidate species for measuring PV that can be brought in the gas phase in a controlled way. This allowed us to report in this work the first high-resolution spectroscopic investigations of the vibrational spectrum of this molecule.

Furthermore, the scaling of the vibrational PV shifts with atomic number $Z$ was investigated by comparison with \ce{Os(acac)3}, where osmium is the heavier homologue of ruthenium. Here, we find enhancements of the PV shifts that exceed the prediction following from the naive $Z^5$ scaling. 
This is especially auspicious for the experiment -- the investigated modes that fall in the accessible laser windows are enhanced over 40-fold to the $10^{-14}$ fractional shift regime, making detection feasible within the current experimental sensitivity. \ce{Os(acac)3} can be synthesised \cite{dallmann1998darstellung} and the next experimental steps are to investigate its enantiomeric resolution and its stability upon heating for bringing it into the gas phase for the high-resolution spectroscopic studies. 
The favourable findings of our theory-experiment investigations pave the way towards first detection of PV effects in vibrational spectra of chiral molecules at LPL.

Moreover, $^{99}$Ru, $^{101}$Ru, $^{187}$Os and $^{189}$Os nuclei are NMR active and thus the \ce{Ru(acac)3} and \ce{Os(acac)3} complexes potentially open up a possibility for gas phase NMR measurements.

\emph{Acknowledgements. -- }
The authors thank the Center for Information Technology of the University of Groningen for their support and for providing access to the Peregrine high performance computing cluster, and are grateful to Stefan Knecht and Charles Desfrançois for useful discussions.
LFP acknowledges the support from the Slovak Research and Development Agency (APVV-20-0098, APVV-20-0127) and the Scientific Grant Agency of the Slovak Republic (1/0777/19). The authors also acknowledge financial
support from the ANR project PVCM (Grant No. ANR15-CE30-0005-01) and Ile-de-France region (DIM-NanoK).



\bibliography{references}

\begin{thebibliography}{72}%
\makeatletter
\providecommand \@ifxundefined [1]{%
 \@ifx{#1\undefined}
}%
\providecommand \@ifnum [1]{%
 \ifnum #1\expandafter \@firstoftwo
 \else \expandafter \@secondoftwo
 \fi
}%
\providecommand \@ifx [1]{%
 \ifx #1\expandafter \@firstoftwo
 \else \expandafter \@secondoftwo
 \fi
}%
\providecommand \natexlab [1]{#1}%
\providecommand \enquote  [1]{``#1''}%
\providecommand \bibnamefont  [1]{#1}%
\providecommand \bibfnamefont [1]{#1}%
\providecommand \citenamefont [1]{#1}%
\providecommand \href@noop [0]{\@secondoftwo}%
\providecommand \href [0]{\begingroup \@sanitize@url \@href}%
\providecommand \@href[1]{\@@startlink{#1}\@@href}%
\providecommand \@@href[1]{\endgroup#1\@@endlink}%
\providecommand \@sanitize@url [0]{\catcode `\\12\catcode `\$12\catcode
  `\&12\catcode `\#12\catcode `\^12\catcode `\_12\catcode `\%12\relax}%
\providecommand \@@startlink[1]{}%
\providecommand \@@endlink[0]{}%
\providecommand \url  [0]{\begingroup\@sanitize@url \@url }%
\providecommand \@url [1]{\endgroup\@href {#1}{\urlprefix }}%
\providecommand \urlprefix  [0]{URL }%
\providecommand \Eprint [0]{\href }%
\providecommand \doibase [0]{https://doi.org/}%
\providecommand \selectlanguage [0]{\@gobble}%
\providecommand \bibinfo  [0]{\@secondoftwo}%
\providecommand \bibfield  [0]{\@secondoftwo}%
\providecommand \translation [1]{[#1]}%
\providecommand \BibitemOpen [0]{}%
\providecommand \bibitemStop [0]{}%
\providecommand \bibitemNoStop [0]{.\EOS\space}%
\providecommand \EOS [0]{\spacefactor3000\relax}%
\providecommand \BibitemShut  [1]{\csname bibitem#1\endcsname}%
\let\auto@bib@innerbib\@empty
\bibitem [{\citenamefont {Lee}\ and\ \citenamefont {Yang}(1956)}]{leeyang}%
  \BibitemOpen
  \bibfield  {author} {\bibinfo {author} {\bibfnamefont {T.}~\bibnamefont
  {Lee}}\ and\ \bibinfo {author} {\bibfnamefont {C.}~\bibnamefont {Yang}},\
  }\bibfield  {title} {\bibinfo {title} {Question of parity conservation in
  weak interactions},\ }\href {https://doi.org/10.1103/PhysRev.104.254}
  {\bibfield  {journal} {\bibinfo  {journal} {Phys. Rev.}\ }\textbf {\bibinfo
  {volume} {104}},\ \bibinfo {pages} {254} (\bibinfo {year}
  {1956})}\BibitemShut {NoStop}%
\bibitem [{\citenamefont {Wu}\ \emph {et~al.}(1957)\citenamefont {Wu},
  \citenamefont {Ambler}, \citenamefont {Hayward}, \citenamefont {Hoppes},\
  and\ \citenamefont {Hudson}}]{wu1957experimental}%
  \BibitemOpen
  \bibfield  {author} {\bibinfo {author} {\bibfnamefont {C.-S.}\ \bibnamefont
  {Wu}}, \bibinfo {author} {\bibfnamefont {E.}~\bibnamefont {Ambler}}, \bibinfo
  {author} {\bibfnamefont {R.}~\bibnamefont {Hayward}}, \bibinfo {author}
  {\bibfnamefont {D.}~\bibnamefont {Hoppes}},\ and\ \bibinfo {author}
  {\bibfnamefont {R.}~\bibnamefont {Hudson}},\ }\bibfield  {title} {\bibinfo
  {title} {Experimental test of parity conservation in beta decay},\
  }\href@noop {} {\bibfield  {journal} {\bibinfo  {journal} {Phys. Rev.}\
  }\textbf {\bibinfo {volume} {105}},\ \bibinfo {pages} {1413} (\bibinfo {year}
  {1957})}\BibitemShut {NoStop}%
\bibitem [{\citenamefont {Garwin}\ \emph {et~al.}(1957)\citenamefont {Garwin},
  \citenamefont {Lederman},\ and\ \citenamefont {Weinrich}}]{PhysRev.105.1415}%
  \BibitemOpen
  \bibfield  {author} {\bibinfo {author} {\bibfnamefont {R.~L.}\ \bibnamefont
  {Garwin}}, \bibinfo {author} {\bibfnamefont {L.~M.}\ \bibnamefont
  {Lederman}},\ and\ \bibinfo {author} {\bibfnamefont {M.}~\bibnamefont
  {Weinrich}},\ }\bibfield  {title} {\bibinfo {title} {Observations of the
  failure of conservation of parity and charge conjugation in meson decays: the
  magnetic moment of the free muon},\ }\href
  {https://doi.org/10.1103/PhysRev.105.1415} {\bibfield  {journal} {\bibinfo
  {journal} {Phys. Rev.}\ }\textbf {\bibinfo {volume} {105}},\ \bibinfo {pages}
  {1415} (\bibinfo {year} {1957})}\BibitemShut {NoStop}%
\bibitem [{\citenamefont {Barkov}\ and\ \citenamefont
  {Zolotarev}(1978)}]{barko1978measurement}%
  \BibitemOpen
  \bibfield  {author} {\bibinfo {author} {\bibfnamefont {L.}~\bibnamefont
  {Barkov}}\ and\ \bibinfo {author} {\bibfnamefont {M.}~\bibnamefont
  {Zolotarev}},\ }\bibfield  {title} {\bibinfo {title} {Measurement of optical
  activity of bismuth vapor},\ }\href@noop {} {\bibfield  {journal} {\bibinfo
  {journal} {JETP Lett.}\ }\textbf {\bibinfo {volume} {28}},\ \bibinfo {pages}
  {503} (\bibinfo {year} {1978})}\BibitemShut {NoStop}%
\bibitem [{\citenamefont {Bouchiat}\ \emph {et~al.}(1982)\citenamefont
  {Bouchiat}, \citenamefont {Guena}, \citenamefont {Hunter},\ and\
  \citenamefont {Pottier}}]{bouchiat1982observation}%
  \BibitemOpen
  \bibfield  {author} {\bibinfo {author} {\bibfnamefont {M.}~\bibnamefont
  {Bouchiat}}, \bibinfo {author} {\bibfnamefont {J.}~\bibnamefont {Guena}},
  \bibinfo {author} {\bibfnamefont {L.}~\bibnamefont {Hunter}},\ and\ \bibinfo
  {author} {\bibfnamefont {L.}~\bibnamefont {Pottier}},\ }\bibfield  {title}
  {\bibinfo {title} {Observation of a parity violation in cesium},\ }\href@noop
  {} {\bibfield  {journal} {\bibinfo  {journal} {Phys. Lett. B}\ }\textbf
  {\bibinfo {volume} {117}},\ \bibinfo {pages} {358} (\bibinfo {year}
  {1982})}\BibitemShut {NoStop}%
\bibitem [{\citenamefont {Gilbert}\ \emph {et~al.}(1985)\citenamefont
  {Gilbert}, \citenamefont {Noecker}, \citenamefont {Watts},\ and\
  \citenamefont {Wieman}}]{gilbert1985measurement}%
  \BibitemOpen
  \bibfield  {author} {\bibinfo {author} {\bibfnamefont {S.}~\bibnamefont
  {Gilbert}}, \bibinfo {author} {\bibfnamefont {M.}~\bibnamefont {Noecker}},
  \bibinfo {author} {\bibfnamefont {R.}~\bibnamefont {Watts}},\ and\ \bibinfo
  {author} {\bibfnamefont {C.}~\bibnamefont {Wieman}},\ }\bibfield  {title}
  {\bibinfo {title} {Measurement of parity nonconservation in atomic cesium},\
  }\href@noop {} {\bibfield  {journal} {\bibinfo  {journal} {Phys. Rev. Lett.}\
  }\textbf {\bibinfo {volume} {55}},\ \bibinfo {pages} {2680} (\bibinfo {year}
  {1985})}\BibitemShut {NoStop}%
\bibitem [{\citenamefont {Drell}\ and\ \citenamefont
  {Commins}(1984)}]{drell1984parity}%
  \BibitemOpen
  \bibfield  {author} {\bibinfo {author} {\bibfnamefont {P.}~\bibnamefont
  {Drell}}\ and\ \bibinfo {author} {\bibfnamefont {E.}~\bibnamefont
  {Commins}},\ }\bibfield  {title} {\bibinfo {title} {Parity nonconservation in
  atomic thallium},\ }\href@noop {} {\bibfield  {journal} {\bibinfo  {journal}
  {Phys. Rev. Lett.}\ }\textbf {\bibinfo {volume} {53}},\ \bibinfo {pages}
  {968} (\bibinfo {year} {1984})}\BibitemShut {NoStop}%
\bibitem [{\citenamefont {Yamagata}(1966)}]{yamagata1966hypothesis}%
  \BibitemOpen
  \bibfield  {author} {\bibinfo {author} {\bibfnamefont {Y.}~\bibnamefont
  {Yamagata}},\ }\bibfield  {title} {\bibinfo {title} {A hypothesis for the
  asymmetric appearance of biomolecules on {Earth}},\ }\href@noop {} {\bibfield
   {journal} {\bibinfo  {journal} {J. Theor. Biol.}\ }\textbf {\bibinfo
  {volume} {11}},\ \bibinfo {pages} {495} (\bibinfo {year} {1966})}\BibitemShut
  {NoStop}%
\bibitem [{\citenamefont {Mason}(1984)}]{mason1984origins}%
  \BibitemOpen
  \bibfield  {author} {\bibinfo {author} {\bibfnamefont {S.}~\bibnamefont
  {Mason}},\ }\bibfield  {title} {\bibinfo {title} {Origins of biomolecular
  handedness},\ }\href@noop {} {\bibfield  {journal} {\bibinfo  {journal}
  {Nature}\ }\textbf {\bibinfo {volume} {311}},\ \bibinfo {pages} {19}
  (\bibinfo {year} {1984})}\BibitemShut {NoStop}%
\bibitem [{\citenamefont {Tranter}(1985)}]{tranter1985parity}%
  \BibitemOpen
  \bibfield  {author} {\bibinfo {author} {\bibfnamefont {G.}~\bibnamefont
  {Tranter}},\ }\bibfield  {title} {\bibinfo {title} {Parity-violating energy
  differences of chiral minerals and the origin of biomolecular
  homochirality},\ }\href@noop {} {\bibfield  {journal} {\bibinfo  {journal}
  {Nature}\ }\textbf {\bibinfo {volume} {318}},\ \bibinfo {pages} {172}
  (\bibinfo {year} {1985})}\BibitemShut {NoStop}%
\bibitem [{\citenamefont {Gaul}\ \emph {et~al.}(2020)\citenamefont {Gaul},
  \citenamefont {Kozlov}, \citenamefont {Isaev},\ and\ \citenamefont
  {Berger}}]{gaul2020chiral}%
  \BibitemOpen
  \bibfield  {author} {\bibinfo {author} {\bibfnamefont {K.}~\bibnamefont
  {Gaul}}, \bibinfo {author} {\bibfnamefont {M.~G.}\ \bibnamefont {Kozlov}},
  \bibinfo {author} {\bibfnamefont {T.~A.}\ \bibnamefont {Isaev}},\ and\
  \bibinfo {author} {\bibfnamefont {R.}~\bibnamefont {Berger}},\ }\bibfield
  {title} {\bibinfo {title} {Chiral molecules as sensitive probes for direct
  detection of p-odd cosmic fields},\ }\href@noop {} {\bibfield  {journal}
  {\bibinfo  {journal} {Phys. Rev. Lett.}\ }\textbf {\bibinfo {volume} {125}},\
  \bibinfo {pages} {123004} (\bibinfo {year} {2020})}\BibitemShut {NoStop}%
\bibitem [{\citenamefont {Barra}\ \emph {et~al.}(1986)\citenamefont {Barra},
  \citenamefont {Robert},\ and\ \citenamefont {Wiesenfeld}}]{barra1986parity}%
  \BibitemOpen
  \bibfield  {author} {\bibinfo {author} {\bibfnamefont {A.}~\bibnamefont
  {Barra}}, \bibinfo {author} {\bibfnamefont {J.}~\bibnamefont {Robert}},\ and\
  \bibinfo {author} {\bibfnamefont {L.}~\bibnamefont {Wiesenfeld}},\ }\bibfield
   {title} {\bibinfo {title} {Parity non-conservation and {NMR} observables.
  {Calculation of Tl} resonance frequency differences in enantiomers},\
  }\href@noop {} {\bibfield  {journal} {\bibinfo  {journal} {Phys. Lett. A}\
  }\textbf {\bibinfo {volume} {115}},\ \bibinfo {pages} {443} (\bibinfo {year}
  {1986})}\BibitemShut {NoStop}%
\bibitem [{\citenamefont {Robert}\ and\ \citenamefont
  {Barra}(2001)}]{robert2001nmr}%
  \BibitemOpen
  \bibfield  {author} {\bibinfo {author} {\bibfnamefont {J.-B.}\ \bibnamefont
  {Robert}}\ and\ \bibinfo {author} {\bibfnamefont {A.}~\bibnamefont {Barra}},\
  }\bibfield  {title} {\bibinfo {title} {{NMR} and parity nonconservation.
  {Experimental} requirements to observe a difference between enantiomer
  signals},\ }\href@noop {} {\bibfield  {journal} {\bibinfo  {journal}
  {Chirality}\ }\textbf {\bibinfo {volume} {13}},\ \bibinfo {pages} {699}
  (\bibinfo {year} {2001})}\BibitemShut {NoStop}%
\bibitem [{\citenamefont {Bast}\ \emph {et~al.}(2006)\citenamefont {Bast},
  \citenamefont {Schwerdtfeger},\ and\ \citenamefont {Saue}}]{bast2006parity}%
  \BibitemOpen
  \bibfield  {author} {\bibinfo {author} {\bibfnamefont {R.}~\bibnamefont
  {Bast}}, \bibinfo {author} {\bibfnamefont {P.}~\bibnamefont
  {Schwerdtfeger}},\ and\ \bibinfo {author} {\bibfnamefont {T.}~\bibnamefont
  {Saue}},\ }\bibfield  {title} {\bibinfo {title} {Parity nonconservation
  contribution to the nuclear magnetic resonance shielding constants of chiral
  molecules: A four-component relativistic study},\ }\href@noop {} {\bibfield
  {journal} {\bibinfo  {journal} {J. Chem. Phys.}\ }\textbf {\bibinfo {volume}
  {125}},\ \bibinfo {pages} {064504} (\bibinfo {year} {2006})}\BibitemShut
  {NoStop}%
\bibitem [{\citenamefont {Eills}\ \emph {et~al.}(2017)\citenamefont {Eills},
  \citenamefont {Blanchard}, \citenamefont {Bougas}, \citenamefont {Kozlov},
  \citenamefont {Pines},\ and\ \citenamefont {Budker}}]{eills2017measuring}%
  \BibitemOpen
  \bibfield  {author} {\bibinfo {author} {\bibfnamefont {J.}~\bibnamefont
  {Eills}}, \bibinfo {author} {\bibfnamefont {J.~W.}\ \bibnamefont
  {Blanchard}}, \bibinfo {author} {\bibfnamefont {L.}~\bibnamefont {Bougas}},
  \bibinfo {author} {\bibfnamefont {M.~G.}\ \bibnamefont {Kozlov}}, \bibinfo
  {author} {\bibfnamefont {A.}~\bibnamefont {Pines}},\ and\ \bibinfo {author}
  {\bibfnamefont {D.}~\bibnamefont {Budker}},\ }\bibfield  {title} {\bibinfo
  {title} {Measuring molecular parity nonconservation using
  nuclear-magnetic-resonance spectroscopy},\ }\href@noop {} {\bibfield
  {journal} {\bibinfo  {journal} {Phys. Rev. A}\ }\textbf {\bibinfo {volume}
  {96}},\ \bibinfo {pages} {042119} (\bibinfo {year} {2017})}\BibitemShut
  {NoStop}%
\bibitem [{\citenamefont {Harris}\ and\ \citenamefont
  {Stodolsky}(1978)}]{harris1978quantum}%
  \BibitemOpen
  \bibfield  {author} {\bibinfo {author} {\bibfnamefont {R.}~\bibnamefont
  {Harris}}\ and\ \bibinfo {author} {\bibfnamefont {L.}~\bibnamefont
  {Stodolsky}},\ }\bibfield  {title} {\bibinfo {title} {Quantum beats in
  optical activity and weak interactions},\ }\href@noop {} {\bibfield
  {journal} {\bibinfo  {journal} {Phys. Lett. B}\ }\textbf {\bibinfo {volume}
  {78}},\ \bibinfo {pages} {313} (\bibinfo {year} {1978})}\BibitemShut
  {NoStop}%
\bibitem [{\citenamefont {Quack}(1986)}]{quack1986measurement}%
  \BibitemOpen
  \bibfield  {author} {\bibinfo {author} {\bibfnamefont {M.}~\bibnamefont
  {Quack}},\ }\bibfield  {title} {\bibinfo {title} {On the measurement of the
  parity violating energy difference between enantiomers},\ }\href@noop {}
  {\bibfield  {journal} {\bibinfo  {journal} {Chem. Phys. Lett.}\ }\textbf
  {\bibinfo {volume} {132}},\ \bibinfo {pages} {147} (\bibinfo {year}
  {1986})}\BibitemShut {NoStop}%
\bibitem [{\citenamefont {Berger}\ and\ \citenamefont
  {Stohner}(2019)}]{berger2019parity}%
  \BibitemOpen
  \bibfield  {author} {\bibinfo {author} {\bibfnamefont {R.}~\bibnamefont
  {Berger}}\ and\ \bibinfo {author} {\bibfnamefont {J.}~\bibnamefont
  {Stohner}},\ }\bibfield  {title} {\bibinfo {title} {Parity violation},\
  }\href@noop {} {\bibfield  {journal} {\bibinfo  {journal} {Wires. Comput.
  Mol. Sci.}\ }\textbf {\bibinfo {volume} {9}},\ \bibinfo {pages} {e1396}
  (\bibinfo {year} {2019})}\BibitemShut {NoStop}%
\bibitem [{\citenamefont {Berger}(2004)}]{pvinbookberger}%
  \BibitemOpen
  \bibfield  {author} {\bibinfo {author} {\bibfnamefont {R.}~\bibnamefont
  {Berger}},\ }\bibfield  {title} {\bibinfo {title} {Parity-violation effects
  in molecules},\ }in\ \href@noop {} {\emph {\bibinfo {booktitle} {Relativistic
  Electronic Structure Theory: Part 2. Applications}}},\ \bibinfo {editor}
  {edited by\ \bibinfo {editor} {\bibfnamefont {P.}~\bibnamefont
  {Schwerdtfeger}}}\ (\bibinfo  {publisher} {Elsevier B.V.},\ \bibinfo
  {address} {Amsterdam, the Netherlands},\ \bibinfo {year} {2004})\
  Chap.~\bibinfo {chapter} {4}, pp.\ \bibinfo {pages} {188--288}\BibitemShut
  {NoStop}%
\bibitem [{\citenamefont {Schwerdtfeger}(2010)}]{Schwerdtfeger2010}%
  \BibitemOpen
  \bibfield  {author} {\bibinfo {author} {\bibfnamefont {P.}~\bibnamefont
  {Schwerdtfeger}},\ }\bibfield  {title} {\bibinfo {title} {The search for
  parity violation in chiral molecules},\ }in\ \href
  {https://doi.org/10.1002/9783527633272.ch7} {\emph {\bibinfo {booktitle}
  {Computational Spectroscopy}}}\ (\bibinfo  {publisher} {Wiley-{VCH} Verlag},\
  \bibinfo {year} {2010})\ Chap.~\bibinfo {chapter} {7}, pp.\ \bibinfo {pages}
  {201--221}\BibitemShut {NoStop}%
\bibitem [{\citenamefont {Daussy}\ \emph {et~al.}(1999)\citenamefont {Daussy},
  \citenamefont {Marrel}, \citenamefont {Amy-Klein}, \citenamefont {Nguyen},
  \citenamefont {Bord{\'e}},\ and\ \citenamefont
  {Chardonnet}}]{daussy1999limit}%
  \BibitemOpen
  \bibfield  {author} {\bibinfo {author} {\bibfnamefont {C.}~\bibnamefont
  {Daussy}}, \bibinfo {author} {\bibfnamefont {T.}~\bibnamefont {Marrel}},
  \bibinfo {author} {\bibfnamefont {A.}~\bibnamefont {Amy-Klein}}, \bibinfo
  {author} {\bibfnamefont {C.}~\bibnamefont {Nguyen}}, \bibinfo {author}
  {\bibfnamefont {C.~J.}\ \bibnamefont {Bord{\'e}}},\ and\ \bibinfo {author}
  {\bibfnamefont {C.}~\bibnamefont {Chardonnet}},\ }\bibfield  {title}
  {\bibinfo {title} {Limit on the parity nonconserving energy difference
  between the enantiomers of a chiral molecule by laser spectroscopy},\
  }\href@noop {} {\bibfield  {journal} {\bibinfo  {journal} {Phys. Rev. Lett.}\
  }\textbf {\bibinfo {volume} {83}},\ \bibinfo {pages} {1554} (\bibinfo {year}
  {1999})}\BibitemShut {NoStop}%
\bibitem [{\citenamefont {Ziskind}\ \emph {et~al.}(2002)\citenamefont
  {Ziskind}, \citenamefont {Daussy}, \citenamefont {Marrel},\ and\
  \citenamefont {Chardonnet}}]{ziskind2002improved}%
  \BibitemOpen
  \bibfield  {author} {\bibinfo {author} {\bibfnamefont {M.}~\bibnamefont
  {Ziskind}}, \bibinfo {author} {\bibfnamefont {C.}~\bibnamefont {Daussy}},
  \bibinfo {author} {\bibfnamefont {T.}~\bibnamefont {Marrel}},\ and\ \bibinfo
  {author} {\bibfnamefont {C.}~\bibnamefont {Chardonnet}},\ }\bibfield  {title}
  {\bibinfo {title} {Improved sensitivity in the search for a parity-violating
  energy difference in the vibrational spectrum of the enantiomers of
  {CHFClBr}},\ }\href@noop {} {\bibfield  {journal} {\bibinfo  {journal} {Eur.
  Phys. J. D}\ }\textbf {\bibinfo {volume} {20}},\ \bibinfo {pages} {219}
  (\bibinfo {year} {2002})}\BibitemShut {NoStop}%
\bibitem [{\citenamefont {Darqui{\'e}}\ \emph {et~al.}(2010)\citenamefont
  {Darqui{\'e}}, \citenamefont {Stoeffler}, \citenamefont {Shelkovnikov},
  \citenamefont {Daussy}, \citenamefont {Amy-Klein}, \citenamefont
  {Chardonnet}, \citenamefont {Zrig}, \citenamefont {Guy}, \citenamefont
  {Crassous}, \citenamefont {Soulard}, \citenamefont {Asselin}, \citenamefont
  {Huet}, \citenamefont {Schwerdtfeger}, \citenamefont {Bast},\ and\
  \citenamefont {Saue}}]{darquie2010progress}%
  \BibitemOpen
  \bibfield  {author} {\bibinfo {author} {\bibfnamefont {B.}~\bibnamefont
  {Darqui{\'e}}}, \bibinfo {author} {\bibfnamefont {C.}~\bibnamefont
  {Stoeffler}}, \bibinfo {author} {\bibfnamefont {A.}~\bibnamefont
  {Shelkovnikov}}, \bibinfo {author} {\bibfnamefont {C.}~\bibnamefont
  {Daussy}}, \bibinfo {author} {\bibfnamefont {A.}~\bibnamefont {Amy-Klein}},
  \bibinfo {author} {\bibfnamefont {C.}~\bibnamefont {Chardonnet}}, \bibinfo
  {author} {\bibfnamefont {S.}~\bibnamefont {Zrig}}, \bibinfo {author}
  {\bibfnamefont {L.}~\bibnamefont {Guy}}, \bibinfo {author} {\bibfnamefont
  {J.}~\bibnamefont {Crassous}}, \bibinfo {author} {\bibfnamefont
  {P.}~\bibnamefont {Soulard}}, \bibinfo {author} {\bibfnamefont
  {P.}~\bibnamefont {Asselin}}, \bibinfo {author} {\bibfnamefont
  {T.}~\bibnamefont {Huet}}, \bibinfo {author} {\bibfnamefont {P.}~\bibnamefont
  {Schwerdtfeger}}, \bibinfo {author} {\bibfnamefont {R.}~\bibnamefont
  {Bast}},\ and\ \bibinfo {author} {\bibfnamefont {T.}~\bibnamefont {Saue}},\
  }\bibfield  {title} {\bibinfo {title} {Progress toward the first observation
  of parity violation in chiral molecules by high-resolution laser
  spectroscopy},\ }\href@noop {} {\bibfield  {journal} {\bibinfo  {journal}
  {Chirality}\ }\textbf {\bibinfo {volume} {22}},\ \bibinfo {pages} {870}
  (\bibinfo {year} {2010})}\BibitemShut {NoStop}%
\bibitem [{\citenamefont {Tokunaga}\ \emph {et~al.}(2013)\citenamefont
  {Tokunaga}, \citenamefont {Stoeffler}, \citenamefont {Auguste}, \citenamefont
  {Shelkovnikov}, \citenamefont {Daussy}, \citenamefont {Amy-Klein},
  \citenamefont {Chardonnet},\ and\ \citenamefont
  {Darqui{\'e}}}]{tokunaga2013probing}%
  \BibitemOpen
  \bibfield  {author} {\bibinfo {author} {\bibfnamefont {S.~K.}\ \bibnamefont
  {Tokunaga}}, \bibinfo {author} {\bibfnamefont {C.}~\bibnamefont {Stoeffler}},
  \bibinfo {author} {\bibfnamefont {F.}~\bibnamefont {Auguste}}, \bibinfo
  {author} {\bibfnamefont {A.}~\bibnamefont {Shelkovnikov}}, \bibinfo {author}
  {\bibfnamefont {C.}~\bibnamefont {Daussy}}, \bibinfo {author} {\bibfnamefont
  {A.}~\bibnamefont {Amy-Klein}}, \bibinfo {author} {\bibfnamefont
  {C.}~\bibnamefont {Chardonnet}},\ and\ \bibinfo {author} {\bibfnamefont
  {B.}~\bibnamefont {Darqui{\'e}}},\ }\bibfield  {title} {\bibinfo {title}
  {Probing weak force-induced parity violation by high-resolution mid-infrared
  molecular spectroscopy},\ }\href@noop {} {\bibfield  {journal} {\bibinfo
  {journal} {Mol. Phys.}\ }\textbf {\bibinfo {volume} {111}},\ \bibinfo {pages}
  {2363} (\bibinfo {year} {2013})}\BibitemShut {NoStop}%
\bibitem [{\citenamefont {Cournol}\ \emph {et~al.}(2019)\citenamefont
  {Cournol}, \citenamefont {Manceau}, \citenamefont {Pierens}, \citenamefont
  {Lecordier}, \citenamefont {Tran}, \citenamefont {Santagata}, \citenamefont
  {Argence}, \citenamefont {Goncharov}, \citenamefont {Lopez}, \citenamefont
  {Abgrall}, \citenamefont {Le~Coq}, \citenamefont {Le~Targat}, \citenamefont
  {Álvarez Martinez}, \citenamefont {Lee}, \citenamefont {Xu}, \citenamefont
  {Pottie}, \citenamefont {Hendricks}, \citenamefont {Wall}, \citenamefont
  {Bieniewska}, \citenamefont {Sauer}, \citenamefont {Tarbutt}, \citenamefont
  {Amy-Klein}, \citenamefont {Tokunaga},\ and\ \citenamefont
  {Darquié}}]{cournol2019new}%
  \BibitemOpen
  \bibfield  {author} {\bibinfo {author} {\bibfnamefont {A.}~\bibnamefont
  {Cournol}}, \bibinfo {author} {\bibfnamefont {M.}~\bibnamefont {Manceau}},
  \bibinfo {author} {\bibfnamefont {M.}~\bibnamefont {Pierens}}, \bibinfo
  {author} {\bibfnamefont {L.}~\bibnamefont {Lecordier}}, \bibinfo {author}
  {\bibfnamefont {D.}~\bibnamefont {Tran}}, \bibinfo {author} {\bibfnamefont
  {R.}~\bibnamefont {Santagata}}, \bibinfo {author} {\bibfnamefont
  {B.}~\bibnamefont {Argence}}, \bibinfo {author} {\bibfnamefont
  {A.}~\bibnamefont {Goncharov}}, \bibinfo {author} {\bibfnamefont
  {O.}~\bibnamefont {Lopez}}, \bibinfo {author} {\bibfnamefont
  {M.}~\bibnamefont {Abgrall}}, \bibinfo {author} {\bibfnamefont
  {Y.}~\bibnamefont {Le~Coq}}, \bibinfo {author} {\bibfnamefont
  {R.}~\bibnamefont {Le~Targat}}, \bibinfo {author} {\bibfnamefont
  {H.}~\bibnamefont {Álvarez Martinez}}, \bibinfo {author} {\bibfnamefont
  {W.}~\bibnamefont {Lee}}, \bibinfo {author} {\bibfnamefont {D.}~\bibnamefont
  {Xu}}, \bibinfo {author} {\bibfnamefont {P.-E.}\ \bibnamefont {Pottie}},
  \bibinfo {author} {\bibfnamefont {R.}~\bibnamefont {Hendricks}}, \bibinfo
  {author} {\bibfnamefont {T.}~\bibnamefont {Wall}}, \bibinfo {author}
  {\bibfnamefont {J.}~\bibnamefont {Bieniewska}}, \bibinfo {author}
  {\bibfnamefont {B.}~\bibnamefont {Sauer}}, \bibinfo {author} {\bibfnamefont
  {M.}~\bibnamefont {Tarbutt}}, \bibinfo {author} {\bibfnamefont
  {A.}~\bibnamefont {Amy-Klein}}, \bibinfo {author} {\bibfnamefont
  {S.}~\bibnamefont {Tokunaga}},\ and\ \bibinfo {author} {\bibfnamefont
  {B.}~\bibnamefont {Darquié}},\ }\bibfield  {title} {\bibinfo {title} {A new
  experiment to test parity symmetry in cold chiral molecules using vibrational
  spectroscopy},\ }\href@noop {} {\bibfield  {journal} {\bibinfo  {journal}
  {Quantum Electronics}\ }\textbf {\bibinfo {volume} {49}},\ \bibinfo {pages}
  {288} (\bibinfo {year} {2019})}\BibitemShut {NoStop}%
\bibitem [{\citenamefont {Kompanets}\ \emph {et~al.}(1976)\citenamefont
  {Kompanets}, \citenamefont {Kukudzhanov}, \citenamefont {Letokhov},\ and\
  \citenamefont {Gervits}}]{kompanets1976narrow}%
  \BibitemOpen
  \bibfield  {author} {\bibinfo {author} {\bibfnamefont {O.}~\bibnamefont
  {Kompanets}}, \bibinfo {author} {\bibfnamefont {A.}~\bibnamefont
  {Kukudzhanov}}, \bibinfo {author} {\bibfnamefont {V.}~\bibnamefont
  {Letokhov}},\ and\ \bibinfo {author} {\bibfnamefont {L.}~\bibnamefont
  {Gervits}},\ }\bibfield  {title} {\bibinfo {title} {Narrow resonances of
  saturated absorption of the asymmetrical molecule {CHFCIBr} and the
  possibility of weak current detection in molecular physics},\ }\href@noop {}
  {\bibfield  {journal} {\bibinfo  {journal} {Opt. Commun.}\ }\textbf {\bibinfo
  {volume} {19}},\ \bibinfo {pages} {414} (\bibinfo {year} {1976})}\BibitemShut
  {NoStop}%
\bibitem [{\citenamefont {Schwerdtfeger}\ \emph {et~al.}(2002)\citenamefont
  {Schwerdtfeger}, \citenamefont {Laerdahl},\ and\ \citenamefont
  {Chardonnet}}]{Schwerdtfeger2002}%
  \BibitemOpen
  \bibfield  {author} {\bibinfo {author} {\bibfnamefont {P.}~\bibnamefont
  {Schwerdtfeger}}, \bibinfo {author} {\bibfnamefont {J.~K.}\ \bibnamefont
  {Laerdahl}},\ and\ \bibinfo {author} {\bibfnamefont {C.}~\bibnamefont
  {Chardonnet}},\ }\bibfield  {title} {\bibinfo {title} {Calculation of
  parity-violation effects for the {C-F} stretching mode of chiral methyl
  fluorides},\ }\href {https://doi.org/10.1103/PhysRevA.65.042508} {\bibfield
  {journal} {\bibinfo  {journal} {Phys. Rev. A}\ }\textbf {\bibinfo {volume}
  {65}},\ \bibinfo {pages} {042508} (\bibinfo {year} {2002})}\BibitemShut
  {NoStop}%
\bibitem [{\citenamefont {Quack}\ and\ \citenamefont
  {Stohner}(2003)}]{quack2003combined}%
  \BibitemOpen
  \bibfield  {author} {\bibinfo {author} {\bibfnamefont {M.}~\bibnamefont
  {Quack}}\ and\ \bibinfo {author} {\bibfnamefont {J.}~\bibnamefont
  {Stohner}},\ }\bibfield  {title} {\bibinfo {title} {Combined multidimensional
  anharmonic and parity violating effects in {CDBrClF}},\ }\href@noop {}
  {\bibfield  {journal} {\bibinfo  {journal} {J. Chem. Phys.}\ }\textbf
  {\bibinfo {volume} {119}},\ \bibinfo {pages} {11228} (\bibinfo {year}
  {2003})}\BibitemShut {NoStop}%
\bibitem [{\citenamefont {Schwerdtfeger}\ \emph {et~al.}(2005)\citenamefont
  {Schwerdtfeger}, \citenamefont {Saue}, \citenamefont {van Stralen},\ and\
  \citenamefont {Visscher}}]{schwerdtfeger2005relativistic}%
  \BibitemOpen
  \bibfield  {author} {\bibinfo {author} {\bibfnamefont {P.}~\bibnamefont
  {Schwerdtfeger}}, \bibinfo {author} {\bibfnamefont {T.}~\bibnamefont {Saue}},
  \bibinfo {author} {\bibfnamefont {J.}~\bibnamefont {van Stralen}},\ and\
  \bibinfo {author} {\bibfnamefont {L.}~\bibnamefont {Visscher}},\ }\bibfield
  {title} {\bibinfo {title} {Relativistic second-order many-body and
  density-functional theory for the parity-violation contribution to the {C-F}
  stretching mode in {CHFClBr}},\ }\href@noop {} {\bibfield  {journal}
  {\bibinfo  {journal} {Phys. Rev. A}\ }\textbf {\bibinfo {volume} {71}},\
  \bibinfo {pages} {012103} (\bibinfo {year} {2005})}\BibitemShut {NoStop}%
\bibitem [{\citenamefont {Thierfelder}\ \emph {et~al.}(2010)\citenamefont
  {Thierfelder}, \citenamefont {Rauhut},\ and\ \citenamefont
  {Schwerdtfeger}}]{thierfelder2010relativistic}%
  \BibitemOpen
  \bibfield  {author} {\bibinfo {author} {\bibfnamefont {C.}~\bibnamefont
  {Thierfelder}}, \bibinfo {author} {\bibfnamefont {G.}~\bibnamefont
  {Rauhut}},\ and\ \bibinfo {author} {\bibfnamefont {P.}~\bibnamefont
  {Schwerdtfeger}},\ }\bibfield  {title} {\bibinfo {title} {Relativistic
  coupled-cluster study of the parity-violation energy shift of {CHFClBr}},\
  }\href@noop {} {\bibfield  {journal} {\bibinfo  {journal} {Phys. Rev. A}\
  }\textbf {\bibinfo {volume} {81}},\ \bibinfo {pages} {032513} (\bibinfo
  {year} {2010})}\BibitemShut {NoStop}%
\bibitem [{\citenamefont {Rauhut}\ and\ \citenamefont
  {Schwerdtfeger}(2021)}]{RauSch2021}%
  \BibitemOpen
  \bibfield  {author} {\bibinfo {author} {\bibfnamefont {G.}~\bibnamefont
  {Rauhut}}\ and\ \bibinfo {author} {\bibfnamefont {P.}~\bibnamefont
  {Schwerdtfeger}},\ }\bibfield  {title} {\bibinfo {title} {Parity-violation
  effects in the vibrational spectra of {CHFClBr and CDFClBr}},\ }\href
  {https://doi.org/10.1103/PhysRevA.103.042819} {\bibfield  {journal} {\bibinfo
   {journal} {Phys. Rev. A}\ }\textbf {\bibinfo {volume} {103}},\ \bibinfo
  {pages} {042819} (\bibinfo {year} {2021})}\BibitemShut {NoStop}%
\bibitem [{\citenamefont {Soulard}\ \emph {et~al.}(2006)\citenamefont
  {Soulard}, \citenamefont {Asselin}, \citenamefont {Cuisset}, \citenamefont
  {Moreno}, \citenamefont {Huet}, \citenamefont {Petitprez}, \citenamefont
  {Demaison}, \citenamefont {Freedman}, \citenamefont {Cao}, \citenamefont
  {Nafie} \emph {et~al.}}]{soulard2006chlorofluoroiodomethane}%
  \BibitemOpen
  \bibfield  {author} {\bibinfo {author} {\bibfnamefont {P.}~\bibnamefont
  {Soulard}}, \bibinfo {author} {\bibfnamefont {P.}~\bibnamefont {Asselin}},
  \bibinfo {author} {\bibfnamefont {A.}~\bibnamefont {Cuisset}}, \bibinfo
  {author} {\bibfnamefont {J.~R.~A.}\ \bibnamefont {Moreno}}, \bibinfo {author}
  {\bibfnamefont {T.~R.}\ \bibnamefont {Huet}}, \bibinfo {author}
  {\bibfnamefont {D.}~\bibnamefont {Petitprez}}, \bibinfo {author}
  {\bibfnamefont {J.}~\bibnamefont {Demaison}}, \bibinfo {author}
  {\bibfnamefont {T.~B.}\ \bibnamefont {Freedman}}, \bibinfo {author}
  {\bibfnamefont {X.}~\bibnamefont {Cao}}, \bibinfo {author} {\bibfnamefont
  {L.~A.}\ \bibnamefont {Nafie}}, \emph {et~al.},\ }\bibfield  {title}
  {\bibinfo {title} {Chlorofluoroiodomethane as a potential candidate for
  parity violation measurements},\ }\href@noop {} {\bibfield  {journal}
  {\bibinfo  {journal} {Phys. Chem. Chem. Phys.}\ }\textbf {\bibinfo {volume}
  {8}},\ \bibinfo {pages} {79} (\bibinfo {year} {2006})}\BibitemShut {NoStop}%
\bibitem [{\citenamefont {Figgen}\ and\ \citenamefont
  {Schwerdtfeger}(2008)}]{figgen2008seocli}%
  \BibitemOpen
  \bibfield  {author} {\bibinfo {author} {\bibfnamefont {D.}~\bibnamefont
  {Figgen}}\ and\ \bibinfo {author} {\bibfnamefont {P.}~\bibnamefont
  {Schwerdtfeger}},\ }\bibfield  {title} {\bibinfo {title} {{SeOClI}: a
  promising candidate for the detection of parity violation in chiral
  molecules},\ }\href@noop {} {\bibfield  {journal} {\bibinfo  {journal} {Phys.
  Rev. A}\ }\textbf {\bibinfo {volume} {78}},\ \bibinfo {pages} {012511}
  (\bibinfo {year} {2008})}\BibitemShut {NoStop}%
\bibitem [{\citenamefont {Figgen}\ \emph {et~al.}(2010)\citenamefont {Figgen},
  \citenamefont {Koers},\ and\ \citenamefont
  {Schwerdtfeger}}]{figgen2010nwhcli}%
  \BibitemOpen
  \bibfield  {author} {\bibinfo {author} {\bibfnamefont {D.}~\bibnamefont
  {Figgen}}, \bibinfo {author} {\bibfnamefont {A.}~\bibnamefont {Koers}},\ and\
  \bibinfo {author} {\bibfnamefont {P.}~\bibnamefont {Schwerdtfeger}},\
  }\bibfield  {title} {\bibinfo {title} {{NWHClI}: a small and compact chiral
  molecule with large parity-violation effects in the vibrational spectrum},\
  }\href@noop {} {\bibfield  {journal} {\bibinfo  {journal} {Angew. Chem. Int.
  Edit.}\ }\textbf {\bibinfo {volume} {49}},\ \bibinfo {pages} {2941} (\bibinfo
  {year} {2010})}\BibitemShut {NoStop}%
\bibitem [{\citenamefont {Wormit}\ \emph {et~al.}(2014)\citenamefont {Wormit},
  \citenamefont {Olejniczak}, \citenamefont {Deppenmeier}, \citenamefont
  {Borschevsky}, \citenamefont {Saue},\ and\ \citenamefont
  {Schwerdtfeger}}]{wormit2014strong}%
  \BibitemOpen
  \bibfield  {author} {\bibinfo {author} {\bibfnamefont {M.}~\bibnamefont
  {Wormit}}, \bibinfo {author} {\bibfnamefont {M.}~\bibnamefont {Olejniczak}},
  \bibinfo {author} {\bibfnamefont {A.-L.}\ \bibnamefont {Deppenmeier}},
  \bibinfo {author} {\bibfnamefont {A.}~\bibnamefont {Borschevsky}}, \bibinfo
  {author} {\bibfnamefont {T.}~\bibnamefont {Saue}},\ and\ \bibinfo {author}
  {\bibfnamefont {P.}~\bibnamefont {Schwerdtfeger}},\ }\bibfield  {title}
  {\bibinfo {title} {Strong enhancement of parity violation effects in chiral
  uranium compounds},\ }\href@noop {} {\bibfield  {journal} {\bibinfo
  {journal} {Phys. Chem. Chem. Phys.}\ }\textbf {\bibinfo {volume} {16}},\
  \bibinfo {pages} {17043} (\bibinfo {year} {2014})}\BibitemShut {NoStop}%
\bibitem [{\citenamefont {Stoeffler}\ \emph {et~al.}(2011)\citenamefont
  {Stoeffler}, \citenamefont {Darqui{\'e}}, \citenamefont {Shelkovnikov},
  \citenamefont {Daussy}, \citenamefont {Amy-Klein}, \citenamefont
  {Chardonnet}, \citenamefont {Guy}, \citenamefont {Crassous}, \citenamefont
  {Huet}, \citenamefont {Soulard} \emph {et~al.}}]{stoeffler2011high}%
  \BibitemOpen
  \bibfield  {author} {\bibinfo {author} {\bibfnamefont {C.}~\bibnamefont
  {Stoeffler}}, \bibinfo {author} {\bibfnamefont {B.}~\bibnamefont
  {Darqui{\'e}}}, \bibinfo {author} {\bibfnamefont {A.}~\bibnamefont
  {Shelkovnikov}}, \bibinfo {author} {\bibfnamefont {C.}~\bibnamefont
  {Daussy}}, \bibinfo {author} {\bibfnamefont {A.}~\bibnamefont {Amy-Klein}},
  \bibinfo {author} {\bibfnamefont {C.}~\bibnamefont {Chardonnet}}, \bibinfo
  {author} {\bibfnamefont {L.}~\bibnamefont {Guy}}, \bibinfo {author}
  {\bibfnamefont {J.}~\bibnamefont {Crassous}}, \bibinfo {author}
  {\bibfnamefont {T.~R.}\ \bibnamefont {Huet}}, \bibinfo {author}
  {\bibfnamefont {P.}~\bibnamefont {Soulard}}, \emph {et~al.},\ }\bibfield
  {title} {\bibinfo {title} {High resolution spectroscopy of
  methyltrioxorhenium: towards the observation of parity violation in chiral
  molecules},\ }\href@noop {} {\bibfield  {journal} {\bibinfo  {journal} {Phys.
  Chem. Chem. Phys.}\ }\textbf {\bibinfo {volume} {13}},\ \bibinfo {pages}
  {854} (\bibinfo {year} {2011})}\BibitemShut {NoStop}%
\bibitem [{\citenamefont {Saleh}\ \emph {et~al.}(2013)\citenamefont {Saleh},
  \citenamefont {Zrig}, \citenamefont {Roisnel}, \citenamefont {Guy},
  \citenamefont {Bast}, \citenamefont {Saue}, \citenamefont {Darqui{\'e}},\
  and\ \citenamefont {Crassous}}]{saleh2013chiral}%
  \BibitemOpen
  \bibfield  {author} {\bibinfo {author} {\bibfnamefont {N.}~\bibnamefont
  {Saleh}}, \bibinfo {author} {\bibfnamefont {S.}~\bibnamefont {Zrig}},
  \bibinfo {author} {\bibfnamefont {T.}~\bibnamefont {Roisnel}}, \bibinfo
  {author} {\bibfnamefont {L.}~\bibnamefont {Guy}}, \bibinfo {author}
  {\bibfnamefont {R.}~\bibnamefont {Bast}}, \bibinfo {author} {\bibfnamefont
  {T.}~\bibnamefont {Saue}}, \bibinfo {author} {\bibfnamefont {B.}~\bibnamefont
  {Darqui{\'e}}},\ and\ \bibinfo {author} {\bibfnamefont {J.}~\bibnamefont
  {Crassous}},\ }\bibfield  {title} {\bibinfo {title} {A chiral rhenium complex
  with predicted high parity violation effects: synthesis, stereochemical
  characterization by {VCD} spectroscopy and quantum chemical calculations},\
  }\href@noop {} {\bibfield  {journal} {\bibinfo  {journal} {Phys. Chem. Chem.
  Phys.}\ }\textbf {\bibinfo {volume} {15}},\ \bibinfo {pages} {10952}
  (\bibinfo {year} {2013})}\BibitemShut {NoStop}%
\bibitem [{\citenamefont {Saleh}\ \emph {et~al.}(2018)\citenamefont {Saleh},
  \citenamefont {Bast}, \citenamefont {Vanthuyne}, \citenamefont {Roussel},
  \citenamefont {Saue}, \citenamefont {Darqui{\'e}},\ and\ \citenamefont
  {Crassous}}]{saleh2018oxorhenium}%
  \BibitemOpen
  \bibfield  {author} {\bibinfo {author} {\bibfnamefont {N.}~\bibnamefont
  {Saleh}}, \bibinfo {author} {\bibfnamefont {R.}~\bibnamefont {Bast}},
  \bibinfo {author} {\bibfnamefont {N.}~\bibnamefont {Vanthuyne}}, \bibinfo
  {author} {\bibfnamefont {C.}~\bibnamefont {Roussel}}, \bibinfo {author}
  {\bibfnamefont {T.}~\bibnamefont {Saue}}, \bibinfo {author} {\bibfnamefont
  {B.}~\bibnamefont {Darqui{\'e}}},\ and\ \bibinfo {author} {\bibfnamefont
  {J.}~\bibnamefont {Crassous}},\ }\bibfield  {title} {\bibinfo {title} {An
  oxorhenium complex bearing a chiral cyclohexane-1-olato-2-thiolato ligand:
  synthesis, stereochemistry, and theoretical study of parity violation
  vibrational frequency shifts},\ }\href@noop {} {\bibfield  {journal}
  {\bibinfo  {journal} {Chirality}\ }\textbf {\bibinfo {volume} {30}},\
  \bibinfo {pages} {147} (\bibinfo {year} {2018})}\BibitemShut {NoStop}%
\bibitem [{\citenamefont {Zel'dovich}\ \emph {et~al.}(1977)\citenamefont
  {Zel'dovich}, \citenamefont {Saakyan},\ and\ \citenamefont
  {Sobel'man}}]{zeldovich1977jetp}%
  \BibitemOpen
  \bibfield  {author} {\bibinfo {author} {\bibfnamefont {B.}~\bibnamefont
  {Zel'dovich}}, \bibinfo {author} {\bibfnamefont {D.}~\bibnamefont
  {Saakyan}},\ and\ \bibinfo {author} {\bibfnamefont {I.}~\bibnamefont
  {Sobel'man}},\ }\bibfield  {title} {\bibinfo {title} {Energy difference
  between right-hand and left-hand molecules, due to parity nonconservation in
  weak interactions of electrons and nuclei},\ }\href@noop {} {\bibfield
  {journal} {\bibinfo  {journal} {JETP Lett.}\ }\textbf {\bibinfo {volume}
  {25}},\ \bibinfo {pages} {94} (\bibinfo {year} {1977})}\BibitemShut {NoStop}%
\bibitem [{\citenamefont {Bast}\ \emph {et~al.}(2011)\citenamefont {Bast},
  \citenamefont {Koers}, \citenamefont {Gomes}, \citenamefont {Ilia{\v{s}}},
  \citenamefont {Visscher}, \citenamefont {Schwerdtfeger},\ and\ \citenamefont
  {Saue}}]{bast2011analysis}%
  \BibitemOpen
  \bibfield  {author} {\bibinfo {author} {\bibfnamefont {R.}~\bibnamefont
  {Bast}}, \bibinfo {author} {\bibfnamefont {A.}~\bibnamefont {Koers}},
  \bibinfo {author} {\bibfnamefont {A.}~\bibnamefont {Gomes}}, \bibinfo
  {author} {\bibfnamefont {M.}~\bibnamefont {Ilia{\v{s}}}}, \bibinfo {author}
  {\bibfnamefont {L.}~\bibnamefont {Visscher}}, \bibinfo {author}
  {\bibfnamefont {P.}~\bibnamefont {Schwerdtfeger}},\ and\ \bibinfo {author}
  {\bibfnamefont {T.}~\bibnamefont {Saue}},\ }\bibfield  {title} {\bibinfo
  {title} {Analysis of parity violation in chiral molecules},\ }\href@noop {}
  {\bibfield  {journal} {\bibinfo  {journal} {Phys. Chem. Chem. Phys.}\
  }\textbf {\bibinfo {volume} {13}},\ \bibinfo {pages} {864} (\bibinfo {year}
  {2011})}\BibitemShut {NoStop}%
\bibitem [{\citenamefont {Sato}\ \emph {et~al.}(2007)\citenamefont {Sato},
  \citenamefont {Taniguchi}, \citenamefont {Nakahashi}, \citenamefont {Monde},\
  and\ \citenamefont {Yamagishi}}]{sato_effects_2007}%
  \BibitemOpen
  \bibfield  {author} {\bibinfo {author} {\bibfnamefont {H.}~\bibnamefont
  {Sato}}, \bibinfo {author} {\bibfnamefont {T.}~\bibnamefont {Taniguchi}},
  \bibinfo {author} {\bibfnamefont {A.}~\bibnamefont {Nakahashi}}, \bibinfo
  {author} {\bibfnamefont {K.}~\bibnamefont {Monde}},\ and\ \bibinfo {author}
  {\bibfnamefont {A.}~\bibnamefont {Yamagishi}},\ }\bibfield  {title} {\bibinfo
  {title} {Effects of {Central} {Metal} {Ions} on {Vibrational} {Circular}
  {Dichroism} {Spectra} of {Tris}-($\beta$-diketonato)metal({III})
  {Complexes}},\ }\href {https://doi.org/10.1021/ic070300i} {\bibfield
  {journal} {\bibinfo  {journal} {Inorg. Chem.}\ }\textbf {\bibinfo {volume}
  {46}},\ \bibinfo {pages} {6755} (\bibinfo {year} {2007})}\BibitemShut
  {NoStop}%
\bibitem [{\citenamefont {Drake}\ \emph {et~al.}(1983)\citenamefont {Drake},
  \citenamefont {Gould}, \citenamefont {Mason}, \citenamefont {Rosini},\ and\
  \citenamefont {Woodley}}]{drake_optical_1983}%
  \BibitemOpen
  \bibfield  {author} {\bibinfo {author} {\bibfnamefont {A.~F.}\ \bibnamefont
  {Drake}}, \bibinfo {author} {\bibfnamefont {J.~M.}\ \bibnamefont {Gould}},
  \bibinfo {author} {\bibfnamefont {S.~F.}\ \bibnamefont {Mason}}, \bibinfo
  {author} {\bibfnamefont {C.}~\bibnamefont {Rosini}},\ and\ \bibinfo {author}
  {\bibfnamefont {F.~J.}\ \bibnamefont {Woodley}},\ }\bibfield  {title}
  {\bibinfo {title} {The optical resolution of
  tris(pentane-2,4-dionato)metal({III}) complexes},\ }\href
  {https://doi.org/10.1016/S0277-5387(00)87108-9} {\bibfield  {journal}
  {\bibinfo  {journal} {Polyhedron}\ }\textbf {\bibinfo {volume} {2}},\
  \bibinfo {pages} {537} (\bibinfo {year} {1983})}\BibitemShut {NoStop}%
\bibitem [{\citenamefont {Darquié}\ \emph {et~al.}()\citenamefont {Darquié},
  \citenamefont {Saleh}, \citenamefont {Tokunaga}, \citenamefont
  {Srebro-Hooper}, \citenamefont {Ponzi}, \citenamefont {Autschbach},
  \citenamefont {Decleva}, \citenamefont {Garcia}, \citenamefont {Crassous},\
  and\ \citenamefont {Nahon}}]{darquie_pecd_2020}%
  \BibitemOpen
  \bibfield  {author} {\bibinfo {author} {\bibfnamefont {B.}~\bibnamefont
  {Darquié}}, \bibinfo {author} {\bibfnamefont {N.}~\bibnamefont {Saleh}},
  \bibinfo {author} {\bibfnamefont {S.~K.}\ \bibnamefont {Tokunaga}}, \bibinfo
  {author} {\bibfnamefont {M.}~\bibnamefont {Srebro-Hooper}}, \bibinfo {author}
  {\bibfnamefont {A.}~\bibnamefont {Ponzi}}, \bibinfo {author} {\bibfnamefont
  {J.}~\bibnamefont {Autschbach}}, \bibinfo {author} {\bibfnamefont
  {P.}~\bibnamefont {Decleva}}, \bibinfo {author} {\bibfnamefont {G.~A.}\
  \bibnamefont {Garcia}}, \bibinfo {author} {\bibfnamefont {J.}~\bibnamefont
  {Crassous}},\ and\ \bibinfo {author} {\bibfnamefont {L.}~\bibnamefont
  {Nahon}},\ }\bibfield  {title} {\bibinfo {title} {Valence-shell photoelectron
  circular dichroism ruthenium{(III)}-tris-(acetylacetonato) gas-phase
  enantiomers},\ }\href@noop {} {\bibinfo  {journal} {accepted for publication
  in Phys. Chem. Chem. Phys. (2021)}\ }\BibitemShut {NoStop}%
\bibitem [{\citenamefont {Asselin}\ \emph {et~al.}(2017)\citenamefont
  {Asselin}, \citenamefont {Berger}, \citenamefont {Huet}, \citenamefont
  {Margulès}, \citenamefont {Motiyenko}, \citenamefont {Hendricks},
  \citenamefont {Tarbutt}, \citenamefont {Tokunaga},\ and\ \citenamefont
  {Darquié}}]{asselin_characterising_2017}%
  \BibitemOpen
\bibfield  {journal} {  }\bibfield  {author} {\bibinfo {author} {\bibfnamefont
  {P.}~\bibnamefont {Asselin}}, \bibinfo {author} {\bibfnamefont
  {Y.}~\bibnamefont {Berger}}, \bibinfo {author} {\bibfnamefont {T.~R.}\
  \bibnamefont {Huet}}, \bibinfo {author} {\bibfnamefont {L.}~\bibnamefont
  {Margulès}}, \bibinfo {author} {\bibfnamefont {R.}~\bibnamefont
  {Motiyenko}}, \bibinfo {author} {\bibfnamefont {R.~J.}\ \bibnamefont
  {Hendricks}}, \bibinfo {author} {\bibfnamefont {M.~R.}\ \bibnamefont
  {Tarbutt}}, \bibinfo {author} {\bibfnamefont {S.~K.}\ \bibnamefont
  {Tokunaga}},\ and\ \bibinfo {author} {\bibfnamefont {B.}~\bibnamefont
  {Darquié}},\ }\bibfield  {title} {\bibinfo {title} {Characterising molecules
  for fundamental physics: an accurate spectroscopic model of
  methyltrioxorhenium derived from new infrared and millimetre-wave
  measurements},\ }\href {https://doi.org/10.1039/C6CP08724H} {\bibfield
  {journal} {\bibinfo  {journal} {Phys. Chem. Chem. Phys.}\ }\textbf {\bibinfo
  {volume} {19}},\ \bibinfo {pages} {4576} (\bibinfo {year}
  {2017})}\BibitemShut {NoStop}%
\bibitem [{\citenamefont {Tokunaga}\ \emph {et~al.}(2017)\citenamefont
  {Tokunaga}, \citenamefont {Hendricks}, \citenamefont {Tarbutt},\ and\
  \citenamefont {Darquié}}]{Tokunaga2017}%
  \BibitemOpen
  \bibfield  {author} {\bibinfo {author} {\bibfnamefont {S.~K.}\ \bibnamefont
  {Tokunaga}}, \bibinfo {author} {\bibfnamefont {R.~J.}\ \bibnamefont
  {Hendricks}}, \bibinfo {author} {\bibfnamefont {M.~R.}\ \bibnamefont
  {Tarbutt}},\ and\ \bibinfo {author} {\bibfnamefont {B.}~\bibnamefont
  {Darquié}},\ }\bibfield  {title} {\bibinfo {title} {High-resolution
  mid-infrared spectroscopy of buffer-gas-cooled methyltrioxorhenium
  molecules},\ }\href {https://doi.org/10.1088/1367-2630/aa6de4} {\bibfield
  {journal} {\bibinfo  {journal} {N. J. Phys.}\ }\textbf {\bibinfo {volume}
  {19}},\ \bibinfo {pages} {053006} (\bibinfo {year} {2017})}\BibitemShut
  {NoStop}%
\bibitem [{\citenamefont {Dallmann}\ and\ \citenamefont
  {Preetz}(1998)}]{dallmann1998darstellung}%
  \BibitemOpen
  \bibfield  {author} {\bibinfo {author} {\bibfnamefont {K.}~\bibnamefont
  {Dallmann}}\ and\ \bibinfo {author} {\bibfnamefont {W.}~\bibnamefont
  {Preetz}},\ }\bibfield  {title} {\bibinfo {title} {Darstellung,
  kristallstruktur, schwingungsspektren und normalkoordinatenanalyse von
  {[Os(acac)$_3$]}/synthesis, crystal structure, vibrational spectra, and
  normal coordinate analysis of {[Os(acac)$_3$]}},\ }\href@noop {} {\bibfield
  {journal} {\bibinfo  {journal} {Z. Naturforsch. B}\ }\textbf {\bibinfo
  {volume} {53}},\ \bibinfo {pages} {232} (\bibinfo {year} {1998})}\BibitemShut
  {NoStop}%
\bibitem [{\citenamefont {Nahrwold}(2011)}]{nahrwold2011electroweak}%
  \BibitemOpen
  \bibfield  {author} {\bibinfo {author} {\bibfnamefont {S.}~\bibnamefont
  {Nahrwold}},\ }\emph {\bibinfo {title} {Electroweak quantum chemistry: Parity
  violation in spectra of chiral molecules containing heavy atoms}},\
  \href@noop {} {Ph.D. thesis},\ \bibinfo  {school} {Univ.-Bibliothek Frankfurt
  am Main} (\bibinfo {year} {2011})\BibitemShut {NoStop}%
\bibitem [{SI()}]{SI}%
  \BibitemOpen
  \href@noop {} {}\bibinfo {note} {See Supplementary Material [url] for
  experimental details on the synthesis and neon matrix-isolation Fourier
  transform infrared spectroscopy of Ru(acac)$_3$, full computational details
  of the PV shifts calculations, comparison of theoretical and experimental
  vibrational spectra for both Ru(acac)$_3$ and Os(acac)$_3$, an assignment of
  the observed vibrational bands for Ru(acac)$_3$, and an analysis of the
  robustness of the calculations, which includes Refs.
  \cite{Danset2003,churakov_isotopically_1996,hanson2019benchmarking,g16,weigend2005balanced,weigend2006accurate,b3lyp,grimme2010consistent,peterson2007energy,figgen2009energy,DIRAC18,thierfelder2010relativistic,yanai2004new,dyallHv2z,dyallRuv3z,iliavs2007infinite,schimmelpfennig1996amfi,figgen2010nwhcli}.}\BibitemShut
  {Stop}%
\bibitem [{\citenamefont {Noumerov}(1924)}]{noumerov1924method}%
  \BibitemOpen
  \bibfield  {author} {\bibinfo {author} {\bibfnamefont {B.}~\bibnamefont
  {Noumerov}},\ }\bibfield  {title} {\bibinfo {title} {A method of
  extrapolation of perturbations},\ }\href@noop {} {\bibfield  {journal}
  {\bibinfo  {journal} {Mon. Not. R. Astron. Soc.}\ }\textbf {\bibinfo {volume}
  {84}},\ \bibinfo {pages} {592} (\bibinfo {year} {1924})}\BibitemShut
  {NoStop}%
\bibitem [{\citenamefont {Cooley}(1961)}]{cooley1961improved}%
  \BibitemOpen
  \bibfield  {author} {\bibinfo {author} {\bibfnamefont {J.}~\bibnamefont
  {Cooley}},\ }\bibfield  {title} {\bibinfo {title} {An improved eigenvalue
  corrector formula for solving the {Schr{\"o}dinger} equation for central
  fields},\ }\href@noop {} {\bibfield  {journal} {\bibinfo  {journal} {Math.
  Comput.}\ }\textbf {\bibinfo {volume} {15}},\ \bibinfo {pages} {363}
  (\bibinfo {year} {1961})}\BibitemShut {NoStop}%
\bibitem [{\citenamefont {Bast}(2017)}]{radovancode}%
  \BibitemOpen
  \bibfield  {author} {\bibinfo {author} {\bibfnamefont {R.}~\bibnamefont
  {Bast}},\ }\href@noop {} {\bibinfo {title} {Numerov 0.5.0}},\ \bibinfo
  {howpublished} {\url{https://doi.org/10.5281/zenodo.1000406}} (\bibinfo
  {year} {October 2017})\BibitemShut {NoStop}%
\bibitem [{\citenamefont {Chanteau}\ \emph {et~al.}(2013)\citenamefont
  {Chanteau}, \citenamefont {Lopez}, \citenamefont {Zhang}, \citenamefont
  {Nicolodi}, \citenamefont {Argence}, \citenamefont {Auguste}, \citenamefont
  {Abgrall}, \citenamefont {Chardonnet}, \citenamefont {Santarelli},
  \citenamefont {Darquié}, \citenamefont {Le~Coq},\ and\ \citenamefont
  {Amy-Klein}}]{chanteau_mid-infrared_2013}%
  \BibitemOpen
  \bibfield  {author} {\bibinfo {author} {\bibfnamefont {B.}~\bibnamefont
  {Chanteau}}, \bibinfo {author} {\bibfnamefont {O.}~\bibnamefont {Lopez}},
  \bibinfo {author} {\bibfnamefont {W.}~\bibnamefont {Zhang}}, \bibinfo
  {author} {\bibfnamefont {D.}~\bibnamefont {Nicolodi}}, \bibinfo {author}
  {\bibfnamefont {B.}~\bibnamefont {Argence}}, \bibinfo {author} {\bibfnamefont
  {F.}~\bibnamefont {Auguste}}, \bibinfo {author} {\bibfnamefont
  {M.}~\bibnamefont {Abgrall}}, \bibinfo {author} {\bibfnamefont
  {C.}~\bibnamefont {Chardonnet}}, \bibinfo {author} {\bibfnamefont
  {G.}~\bibnamefont {Santarelli}}, \bibinfo {author} {\bibfnamefont
  {B.}~\bibnamefont {Darquié}}, \bibinfo {author} {\bibfnamefont
  {Y.}~\bibnamefont {Le~Coq}},\ and\ \bibinfo {author} {\bibfnamefont
  {A.}~\bibnamefont {Amy-Klein}},\ }\bibfield  {title} {\bibinfo {title}
  {Mid-infrared laser phase-locking to a remote near-infrared frequency
  reference for high-precision molecular spectroscopy},\ }\href@noop {}
  {\bibfield  {journal} {\bibinfo  {journal} {N. J. Phys.}\ }\textbf {\bibinfo
  {volume} {15}},\ \bibinfo {pages} {73003} (\bibinfo {year}
  {2013})}\BibitemShut {NoStop}%
\bibitem [{\citenamefont {Sow}\ \emph {et~al.}(2014)\citenamefont {Sow},
  \citenamefont {Mejri}, \citenamefont {Tokunaga}, \citenamefont {Lopez},
  \citenamefont {Goncharov}, \citenamefont {Argence}, \citenamefont
  {Chardonnet}, \citenamefont {Amy-Klein}, \citenamefont {Daussy},\ and\
  \citenamefont {Darquié}}]{sow_widely_2014}%
  \BibitemOpen
  \bibfield  {author} {\bibinfo {author} {\bibfnamefont {P.~L.~T.}\
  \bibnamefont {Sow}}, \bibinfo {author} {\bibfnamefont {S.}~\bibnamefont
  {Mejri}}, \bibinfo {author} {\bibfnamefont {S.~K.}\ \bibnamefont {Tokunaga}},
  \bibinfo {author} {\bibfnamefont {O.}~\bibnamefont {Lopez}}, \bibinfo
  {author} {\bibfnamefont {A.}~\bibnamefont {Goncharov}}, \bibinfo {author}
  {\bibfnamefont {B.}~\bibnamefont {Argence}}, \bibinfo {author} {\bibfnamefont
  {C.}~\bibnamefont {Chardonnet}}, \bibinfo {author} {\bibfnamefont
  {A.}~\bibnamefont {Amy-Klein}}, \bibinfo {author} {\bibfnamefont
  {C.}~\bibnamefont {Daussy}},\ and\ \bibinfo {author} {\bibfnamefont
  {B.}~\bibnamefont {Darquié}},\ }\bibfield  {title} {\bibinfo {title} {A
  widely tunable 10-$\mu$m quantum cascade laser phase-locked to a
  state-of-the-art mid-infrared reference for precision molecular
  spectroscopy},\ }\href {https://doi.org/10.1063/1.4886120} {\bibfield
  {journal} {\bibinfo  {journal} {App. Phys. Lett.}\ }\textbf {\bibinfo
  {volume} {104}},\ \bibinfo {pages} {264101} (\bibinfo {year}
  {2014})}\BibitemShut {NoStop}%
\bibitem [{\citenamefont {Argence}\ \emph {et~al.}(2015)\citenamefont
  {Argence}, \citenamefont {Chanteau}, \citenamefont {Lopez}, \citenamefont
  {Nicolodi}, \citenamefont {Abgrall}, \citenamefont {Chardonnet},
  \citenamefont {Daussy}, \citenamefont {Darquié}, \citenamefont {Le~Coq},\
  and\ \citenamefont {Amy-Klein}}]{argence_quantum_2015}%
  \BibitemOpen
  \bibfield  {author} {\bibinfo {author} {\bibfnamefont {B.}~\bibnamefont
  {Argence}}, \bibinfo {author} {\bibfnamefont {B.}~\bibnamefont {Chanteau}},
  \bibinfo {author} {\bibfnamefont {O.}~\bibnamefont {Lopez}}, \bibinfo
  {author} {\bibfnamefont {D.}~\bibnamefont {Nicolodi}}, \bibinfo {author}
  {\bibfnamefont {M.}~\bibnamefont {Abgrall}}, \bibinfo {author} {\bibfnamefont
  {C.}~\bibnamefont {Chardonnet}}, \bibinfo {author} {\bibfnamefont
  {C.}~\bibnamefont {Daussy}}, \bibinfo {author} {\bibfnamefont
  {B.}~\bibnamefont {Darquié}}, \bibinfo {author} {\bibfnamefont
  {Y.}~\bibnamefont {Le~Coq}},\ and\ \bibinfo {author} {\bibfnamefont
  {A.}~\bibnamefont {Amy-Klein}},\ }\bibfield  {title} {\bibinfo {title}
  {Quantum cascade laser frequency stabilization at the sub-{Hz} level},\
  }\href {https://doi.org/10.1038/nphoton.2015.93} {\bibfield  {journal}
  {\bibinfo  {journal} {Nature Photonics}\ }\textbf {\bibinfo {volume} {9}},\
  \bibinfo {pages} {456} (\bibinfo {year} {2015})}\BibitemShut {NoStop}%
\bibitem [{\citenamefont {Santagata}\ \emph {et~al.}(2019)\citenamefont
  {Santagata}, \citenamefont {Tran}, \citenamefont {Argence}, \citenamefont
  {Lopez}, \citenamefont {Tokunaga}, \citenamefont {Wiotte}, \citenamefont
  {Mouhamad}, \citenamefont {Goncharov}, \citenamefont {Abgrall}, \citenamefont
  {Le~Coq}, \citenamefont {Alvarez-Martinez}, \citenamefont {Le~Targat},
  \citenamefont {Lee}, \citenamefont {Xu}, \citenamefont {Pottie},
  \citenamefont {Darquié},\ and\ \citenamefont
  {Amy-Klein}}]{santagata_high-precision_2019}%
  \BibitemOpen
  \bibfield  {author} {\bibinfo {author} {\bibfnamefont {R.}~\bibnamefont
  {Santagata}}, \bibinfo {author} {\bibfnamefont {D.~B.~A.}\ \bibnamefont
  {Tran}}, \bibinfo {author} {\bibfnamefont {B.}~\bibnamefont {Argence}},
  \bibinfo {author} {\bibfnamefont {O.}~\bibnamefont {Lopez}}, \bibinfo
  {author} {\bibfnamefont {S.~K.}\ \bibnamefont {Tokunaga}}, \bibinfo {author}
  {\bibfnamefont {F.}~\bibnamefont {Wiotte}}, \bibinfo {author} {\bibfnamefont
  {H.}~\bibnamefont {Mouhamad}}, \bibinfo {author} {\bibfnamefont
  {A.}~\bibnamefont {Goncharov}}, \bibinfo {author} {\bibfnamefont
  {M.}~\bibnamefont {Abgrall}}, \bibinfo {author} {\bibfnamefont
  {Y.}~\bibnamefont {Le~Coq}}, \bibinfo {author} {\bibfnamefont
  {H.}~\bibnamefont {Alvarez-Martinez}}, \bibinfo {author} {\bibfnamefont
  {R.}~\bibnamefont {Le~Targat}}, \bibinfo {author} {\bibfnamefont {W.~K.}\
  \bibnamefont {Lee}}, \bibinfo {author} {\bibfnamefont {D.}~\bibnamefont
  {Xu}}, \bibinfo {author} {\bibfnamefont {P.-E.}\ \bibnamefont {Pottie}},
  \bibinfo {author} {\bibfnamefont {B.}~\bibnamefont {Darquié}},\ and\
  \bibinfo {author} {\bibfnamefont {A.}~\bibnamefont {Amy-Klein}},\ }\bibfield
  {title} {\bibinfo {title} {High-precision methanol spectroscopy with a widely
  tunable {SI}-traceable frequency-comb-based mid-infrared {QCL}},\ }\href
  {https://doi.org/10.1364/OPTICA.6.000411} {\bibfield  {journal} {\bibinfo
  {journal} {Optica}\ }\textbf {\bibinfo {volume} {6}},\ \bibinfo {pages} {411}
  (\bibinfo {year} {2019})}\BibitemShut {NoStop}%
\bibitem [{\citenamefont {Shelkovnikov}\ \emph {et~al.}(2008)\citenamefont
  {Shelkovnikov}, \citenamefont {Butcher}, \citenamefont {Chardonnet},\ and\
  \citenamefont {Amy-Klein}}]{shelkovnikov_stability_2008}%
  \BibitemOpen
  \bibfield  {author} {\bibinfo {author} {\bibfnamefont {A.}~\bibnamefont
  {Shelkovnikov}}, \bibinfo {author} {\bibfnamefont {R.~J.}\ \bibnamefont
  {Butcher}}, \bibinfo {author} {\bibfnamefont {C.}~\bibnamefont
  {Chardonnet}},\ and\ \bibinfo {author} {\bibfnamefont {A.}~\bibnamefont
  {Amy-Klein}},\ }\bibfield  {title} {\bibinfo {title} {Stability of the
  proton-to-electron mass ratio},\ }\href@noop {} {\bibfield  {journal}
  {\bibinfo  {journal} {Phys. Rev. Lett.}\ }\textbf {\bibinfo {volume} {100}},\
  \bibinfo {pages} {150801} (\bibinfo {year} {2008})}\BibitemShut {NoStop}%
\bibitem [{\citenamefont {Danset}\ and\ \citenamefont
  {Manceron}(2003)}]{Danset2003}%
  \BibitemOpen
  \bibfield  {author} {\bibinfo {author} {\bibfnamefont {D.}~\bibnamefont
  {Danset}}\ and\ \bibinfo {author} {\bibfnamefont {L.}~\bibnamefont
  {Manceron}},\ }\bibfield  {title} {\bibinfo {title} {Mid- and near-{IR}
  electronic absorption spectrum of {CoO} isolated in solid neon. {Vibronic}
  data for low-lying electronic states},\ }\href
  {https://doi.org/10.1021/jp0357626} {\bibfield  {journal} {\bibinfo
  {journal} {J. Phys. Chem. A}\ }\textbf {\bibinfo {volume} {107}},\ \bibinfo
  {pages} {11324} (\bibinfo {year} {2003})},\ \Eprint
  {https://arxiv.org/abs/https://doi.org/10.1021/jp0357626}
  {https://doi.org/10.1021/jp0357626} \BibitemShut {NoStop}%
\bibitem [{\citenamefont {Churakov}\ and\ \citenamefont
  {Fuss}(1996)}]{churakov_isotopically_1996}%
  \BibitemOpen
  \bibfield  {author} {\bibinfo {author} {\bibfnamefont {V.}~\bibnamefont
  {Churakov}}\ and\ \bibinfo {author} {\bibfnamefont {W.}~\bibnamefont
  {Fuss}},\ }\bibfield  {title} {\bibinfo {title} {Isotopically selective {IR}
  multiphoton dissociation of 1,3,5-trioxane},\ }\href
  {https://doi.org/10.1007/BF01081126} {\bibfield  {journal} {\bibinfo
  {journal} {Appl. Phys. B}\ }\textbf {\bibinfo {volume} {62}},\ \bibinfo
  {pages} {203} (\bibinfo {year} {1996})}\BibitemShut {NoStop}%
\bibitem [{\citenamefont {Hanson-Heine}(2019)}]{hanson2019benchmarking}%
  \BibitemOpen
  \bibfield  {author} {\bibinfo {author} {\bibfnamefont {M.}~\bibnamefont
  {Hanson-Heine}},\ }\bibfield  {title} {\bibinfo {title} {Benchmarking {DFT-D}
  dispersion corrections for anharmonic vibrational frequencies and harmonic
  scaling factors},\ }\href@noop {} {\bibfield  {journal} {\bibinfo  {journal}
  {J. Phys. Chem. A}\ }\textbf {\bibinfo {volume} {123}},\ \bibinfo {pages}
  {9800} (\bibinfo {year} {2019})}\BibitemShut {NoStop}%
\bibitem [{\citenamefont {Frisch}\ \emph {et~al.}(2016)\citenamefont {Frisch},
  \citenamefont {Trucks}, \citenamefont {Schlegel}, \citenamefont {Scuseria},
  \citenamefont {Robb}, \citenamefont {Cheeseman}, \citenamefont {Scalmani},
  \citenamefont {Barone}, \citenamefont {Petersson}, \citenamefont {Nakatsuji},
  \citenamefont {Li}, \citenamefont {Caricato}, \citenamefont {Marenich},
  \citenamefont {Bloino}, \citenamefont {Janesko}, \citenamefont {Gomperts},
  \citenamefont {Mennucci}, \citenamefont {Hratchian}, \citenamefont {Ortiz},
  \citenamefont {Izmaylov}, \citenamefont {Sonnenberg}, \citenamefont
  {Williams-Young}, \citenamefont {Ding}, \citenamefont {Lipparini},
  \citenamefont {Egidi}, \citenamefont {Goings}, \citenamefont {Peng},
  \citenamefont {Petrone}, \citenamefont {Henderson}, \citenamefont
  {Ranasinghe}, \citenamefont {Zakrzewski}, \citenamefont {Gao}, \citenamefont
  {Rega}, \citenamefont {Zheng}, \citenamefont {Liang}, \citenamefont {Hada},
  \citenamefont {Ehara}, \citenamefont {Toyota}, \citenamefont {Fukuda},
  \citenamefont {Hasegawa}, \citenamefont {Ishida}, \citenamefont {Nakajima},
  \citenamefont {Honda}, \citenamefont {Kitao}, \citenamefont {Nakai},
  \citenamefont {Vreven}, \citenamefont {Throssell}, \citenamefont
  {Montgomery}, \citenamefont {Peralta}, \citenamefont {Ogliaro}, \citenamefont
  {Bearpark}, \citenamefont {Heyd}, \citenamefont {Brothers}, \citenamefont
  {Kudin}, \citenamefont {Staroverov}, \citenamefont {Keith}, \citenamefont
  {Kobayashi}, \citenamefont {Normand}, \citenamefont {Raghavachari},
  \citenamefont {Rendell}, \citenamefont {Burant}, \citenamefont {Iyengar},
  \citenamefont {Tomasi}, \citenamefont {Cossi}, \citenamefont {Millam},
  \citenamefont {Klene}, \citenamefont {Adamo}, \citenamefont {Cammi},
  \citenamefont {Ochterski}, \citenamefont {Martin}, \citenamefont {Morokuma},
  \citenamefont {Farkas}, \citenamefont {Foresman},\ and\ \citenamefont
  {Fox}}]{g16}%
  \BibitemOpen
  \bibfield  {author} {\bibinfo {author} {\bibfnamefont {M.~J.}\ \bibnamefont
  {Frisch}}, \bibinfo {author} {\bibfnamefont {G.~W.}\ \bibnamefont {Trucks}},
  \bibinfo {author} {\bibfnamefont {H.~B.}\ \bibnamefont {Schlegel}}, \bibinfo
  {author} {\bibfnamefont {G.~E.}\ \bibnamefont {Scuseria}}, \bibinfo {author}
  {\bibfnamefont {M.~A.}\ \bibnamefont {Robb}}, \bibinfo {author}
  {\bibfnamefont {J.~R.}\ \bibnamefont {Cheeseman}}, \bibinfo {author}
  {\bibfnamefont {G.}~\bibnamefont {Scalmani}}, \bibinfo {author}
  {\bibfnamefont {V.}~\bibnamefont {Barone}}, \bibinfo {author} {\bibfnamefont
  {G.~A.}\ \bibnamefont {Petersson}}, \bibinfo {author} {\bibfnamefont
  {H.}~\bibnamefont {Nakatsuji}}, \bibinfo {author} {\bibfnamefont
  {X.}~\bibnamefont {Li}}, \bibinfo {author} {\bibfnamefont {M.}~\bibnamefont
  {Caricato}}, \bibinfo {author} {\bibfnamefont {A.~V.}\ \bibnamefont
  {Marenich}}, \bibinfo {author} {\bibfnamefont {J.}~\bibnamefont {Bloino}},
  \bibinfo {author} {\bibfnamefont {B.~G.}\ \bibnamefont {Janesko}}, \bibinfo
  {author} {\bibfnamefont {R.}~\bibnamefont {Gomperts}}, \bibinfo {author}
  {\bibfnamefont {B.}~\bibnamefont {Mennucci}}, \bibinfo {author}
  {\bibfnamefont {H.~P.}\ \bibnamefont {Hratchian}}, \bibinfo {author}
  {\bibfnamefont {J.~V.}\ \bibnamefont {Ortiz}}, \bibinfo {author}
  {\bibfnamefont {A.~F.}\ \bibnamefont {Izmaylov}}, \bibinfo {author}
  {\bibfnamefont {J.~L.}\ \bibnamefont {Sonnenberg}}, \bibinfo {author}
  {\bibfnamefont {D.}~\bibnamefont {Williams-Young}}, \bibinfo {author}
  {\bibfnamefont {F.}~\bibnamefont {Ding}}, \bibinfo {author} {\bibfnamefont
  {F.}~\bibnamefont {Lipparini}}, \bibinfo {author} {\bibfnamefont
  {F.}~\bibnamefont {Egidi}}, \bibinfo {author} {\bibfnamefont
  {J.}~\bibnamefont {Goings}}, \bibinfo {author} {\bibfnamefont
  {B.}~\bibnamefont {Peng}}, \bibinfo {author} {\bibfnamefont {A.}~\bibnamefont
  {Petrone}}, \bibinfo {author} {\bibfnamefont {T.}~\bibnamefont {Henderson}},
  \bibinfo {author} {\bibfnamefont {D.}~\bibnamefont {Ranasinghe}}, \bibinfo
  {author} {\bibfnamefont {V.~G.}\ \bibnamefont {Zakrzewski}}, \bibinfo
  {author} {\bibfnamefont {J.}~\bibnamefont {Gao}}, \bibinfo {author}
  {\bibfnamefont {N.}~\bibnamefont {Rega}}, \bibinfo {author} {\bibfnamefont
  {G.}~\bibnamefont {Zheng}}, \bibinfo {author} {\bibfnamefont
  {W.}~\bibnamefont {Liang}}, \bibinfo {author} {\bibfnamefont
  {M.}~\bibnamefont {Hada}}, \bibinfo {author} {\bibfnamefont {M.}~\bibnamefont
  {Ehara}}, \bibinfo {author} {\bibfnamefont {K.}~\bibnamefont {Toyota}},
  \bibinfo {author} {\bibfnamefont {R.}~\bibnamefont {Fukuda}}, \bibinfo
  {author} {\bibfnamefont {J.}~\bibnamefont {Hasegawa}}, \bibinfo {author}
  {\bibfnamefont {M.}~\bibnamefont {Ishida}}, \bibinfo {author} {\bibfnamefont
  {T.}~\bibnamefont {Nakajima}}, \bibinfo {author} {\bibfnamefont
  {Y.}~\bibnamefont {Honda}}, \bibinfo {author} {\bibfnamefont
  {O.}~\bibnamefont {Kitao}}, \bibinfo {author} {\bibfnamefont
  {H.}~\bibnamefont {Nakai}}, \bibinfo {author} {\bibfnamefont
  {T.}~\bibnamefont {Vreven}}, \bibinfo {author} {\bibfnamefont
  {K.}~\bibnamefont {Throssell}}, \bibinfo {author} {\bibfnamefont {J.~A.}\
  \bibnamefont {Montgomery}, \bibfnamefont {{Jr.}}}, \bibinfo {author}
  {\bibfnamefont {J.~E.}\ \bibnamefont {Peralta}}, \bibinfo {author}
  {\bibfnamefont {F.}~\bibnamefont {Ogliaro}}, \bibinfo {author} {\bibfnamefont
  {M.~J.}\ \bibnamefont {Bearpark}}, \bibinfo {author} {\bibfnamefont {J.~J.}\
  \bibnamefont {Heyd}}, \bibinfo {author} {\bibfnamefont {E.~N.}\ \bibnamefont
  {Brothers}}, \bibinfo {author} {\bibfnamefont {K.~N.}\ \bibnamefont {Kudin}},
  \bibinfo {author} {\bibfnamefont {V.~N.}\ \bibnamefont {Staroverov}},
  \bibinfo {author} {\bibfnamefont {T.~A.}\ \bibnamefont {Keith}}, \bibinfo
  {author} {\bibfnamefont {R.}~\bibnamefont {Kobayashi}}, \bibinfo {author}
  {\bibfnamefont {J.}~\bibnamefont {Normand}}, \bibinfo {author} {\bibfnamefont
  {K.}~\bibnamefont {Raghavachari}}, \bibinfo {author} {\bibfnamefont {A.~P.}\
  \bibnamefont {Rendell}}, \bibinfo {author} {\bibfnamefont {J.~C.}\
  \bibnamefont {Burant}}, \bibinfo {author} {\bibfnamefont {S.~S.}\
  \bibnamefont {Iyengar}}, \bibinfo {author} {\bibfnamefont {J.}~\bibnamefont
  {Tomasi}}, \bibinfo {author} {\bibfnamefont {M.}~\bibnamefont {Cossi}},
  \bibinfo {author} {\bibfnamefont {J.~M.}\ \bibnamefont {Millam}}, \bibinfo
  {author} {\bibfnamefont {M.}~\bibnamefont {Klene}}, \bibinfo {author}
  {\bibfnamefont {C.}~\bibnamefont {Adamo}}, \bibinfo {author} {\bibfnamefont
  {R.}~\bibnamefont {Cammi}}, \bibinfo {author} {\bibfnamefont {J.~W.}\
  \bibnamefont {Ochterski}}, \bibinfo {author} {\bibfnamefont {R.~L.}\
  \bibnamefont {Martin}}, \bibinfo {author} {\bibfnamefont {K.}~\bibnamefont
  {Morokuma}}, \bibinfo {author} {\bibfnamefont {O.}~\bibnamefont {Farkas}},
  \bibinfo {author} {\bibfnamefont {J.~B.}\ \bibnamefont {Foresman}},\ and\
  \bibinfo {author} {\bibfnamefont {D.~J.}\ \bibnamefont {Fox}},\ }\href@noop
  {} {\bibinfo {title} {Gaussian 16 {R}evision {B}.01}} (\bibinfo {year}
  {2016}),\ \bibinfo {note} {{Gaussian Inc. Wallingford CT}}\BibitemShut
  {NoStop}%
\bibitem [{\citenamefont {Weigend}\ and\ \citenamefont
  {Ahlrichs}(2005)}]{weigend2005balanced}%
  \BibitemOpen
  \bibfield  {author} {\bibinfo {author} {\bibfnamefont {F.}~\bibnamefont
  {Weigend}}\ and\ \bibinfo {author} {\bibfnamefont {R.}~\bibnamefont
  {Ahlrichs}},\ }\bibfield  {title} {\bibinfo {title} {Balanced basis sets of
  split valence, triple zeta valence and quadruple zeta valence quality for {H
  to Rn: Design} and assessment of accuracy},\ }\href@noop {} {\bibfield
  {journal} {\bibinfo  {journal} {Phys. Chem. Chem. Phys.}\ }\textbf {\bibinfo
  {volume} {7}},\ \bibinfo {pages} {3297} (\bibinfo {year} {2005})}\BibitemShut
  {NoStop}%
\bibitem [{\citenamefont {Weigend}(2006)}]{weigend2006accurate}%
  \BibitemOpen
  \bibfield  {author} {\bibinfo {author} {\bibfnamefont {F.}~\bibnamefont
  {Weigend}},\ }\bibfield  {title} {\bibinfo {title} {Accurate
  {Coulomb}-fitting basis sets for {H to Rn}},\ }\href@noop {} {\bibfield
  {journal} {\bibinfo  {journal} {Phys. Chem. Chem. Phys.}\ }\textbf {\bibinfo
  {volume} {8}},\ \bibinfo {pages} {1057} (\bibinfo {year} {2006})}\BibitemShut
  {NoStop}%
\bibitem [{\citenamefont {Stephens}\ \emph {et~al.}(1994)\citenamefont
  {Stephens}, \citenamefont {Devlin}, \citenamefont {Chabalowski},\ and\
  \citenamefont {Frisch}}]{b3lyp}%
  \BibitemOpen
  \bibfield  {author} {\bibinfo {author} {\bibfnamefont {P.}~\bibnamefont
  {Stephens}}, \bibinfo {author} {\bibfnamefont {F.}~\bibnamefont {Devlin}},
  \bibinfo {author} {\bibfnamefont {C.}~\bibnamefont {Chabalowski}},\ and\
  \bibinfo {author} {\bibfnamefont {M.}~\bibnamefont {Frisch}},\ }\bibfield
  {title} {\bibinfo {title} {\textit{Ab initio} calculation of vibrational
  absorption and circular dichroism spectra using density functional force
  fields},\ }\href@noop {} {\bibfield  {journal} {\bibinfo  {journal} {J. Phys.
  Chem.}\ }\textbf {\bibinfo {volume} {98}},\ \bibinfo {pages} {11623}
  (\bibinfo {year} {1994})}\BibitemShut {NoStop}%
\bibitem [{\citenamefont {Grimme}\ \emph {et~al.}(2010)\citenamefont {Grimme},
  \citenamefont {Antony}, \citenamefont {Ehrlich},\ and\ \citenamefont
  {Krieg}}]{grimme2010consistent}%
  \BibitemOpen
  \bibfield  {author} {\bibinfo {author} {\bibfnamefont {S.}~\bibnamefont
  {Grimme}}, \bibinfo {author} {\bibfnamefont {J.}~\bibnamefont {Antony}},
  \bibinfo {author} {\bibfnamefont {S.}~\bibnamefont {Ehrlich}},\ and\ \bibinfo
  {author} {\bibfnamefont {H.}~\bibnamefont {Krieg}},\ }\bibfield  {title}
  {\bibinfo {title} {A consistent and accurate \textit{ab initio}
  parametrization of density functional dispersion correction {(DFT-D) for the
  94 elements H-Pu}},\ }\href@noop {} {\bibfield  {journal} {\bibinfo
  {journal} {J. Chem. Phys.}\ }\textbf {\bibinfo {volume} {132}},\ \bibinfo
  {pages} {154104} (\bibinfo {year} {2010})}\BibitemShut {NoStop}%
\bibitem [{\citenamefont {Peterson}\ \emph {et~al.}(2007)\citenamefont
  {Peterson}, \citenamefont {Figgen}, \citenamefont {Dolg},\ and\ \citenamefont
  {Stoll}}]{peterson2007energy}%
  \BibitemOpen
  \bibfield  {author} {\bibinfo {author} {\bibfnamefont {K.}~\bibnamefont
  {Peterson}}, \bibinfo {author} {\bibfnamefont {D.}~\bibnamefont {Figgen}},
  \bibinfo {author} {\bibfnamefont {M.}~\bibnamefont {Dolg}},\ and\ \bibinfo
  {author} {\bibfnamefont {H.}~\bibnamefont {Stoll}},\ }\bibfield  {title}
  {\bibinfo {title} {Energy-consistent relativistic pseudopotentials and
  correlation consistent basis sets for the 4d elements {Y--Pd}},\ }\href@noop
  {} {\bibfield  {journal} {\bibinfo  {journal} {J. Chem. Phys.}\ }\textbf
  {\bibinfo {volume} {126}},\ \bibinfo {pages} {124101} (\bibinfo {year}
  {2007})}\BibitemShut {NoStop}%
\bibitem [{\citenamefont {Figgen}\ \emph {et~al.}(2009)\citenamefont {Figgen},
  \citenamefont {Peterson}, \citenamefont {Dolg},\ and\ \citenamefont
  {Stoll}}]{figgen2009energy}%
  \BibitemOpen
  \bibfield  {author} {\bibinfo {author} {\bibfnamefont {D.}~\bibnamefont
  {Figgen}}, \bibinfo {author} {\bibfnamefont {K.~A.}\ \bibnamefont
  {Peterson}}, \bibinfo {author} {\bibfnamefont {M.}~\bibnamefont {Dolg}},\
  and\ \bibinfo {author} {\bibfnamefont {H.}~\bibnamefont {Stoll}},\ }\bibfield
   {title} {\bibinfo {title} {Energy-consistent pseudopotentials and
  correlation consistent basis sets for the 5$d$ elements {Hf--Pt}},\
  }\href@noop {} {\bibfield  {journal} {\bibinfo  {journal} {J. Chem. Phys.}\
  }\textbf {\bibinfo {volume} {130}},\ \bibinfo {pages} {164108} (\bibinfo
  {year} {2009})}\BibitemShut {NoStop}%
\bibitem [{DIR()}]{DIRAC18}%
  \BibitemOpen
  \href@noop {} {}\bibinfo {note} {{DIRAC}, a relativistic \textit{ab initio}
  electronic structure program, Release {DIRAC18} (2018), written by T.~Saue,
  L.~Visscher, H.~J.~{\relax Aa}.~Jensen, and R.~Bast, with contributions from
  V.~Bakken, K.~G.~Dyall, S.~Dubillard, U.~Ekstr{\"o}m, E.~Eliav,
  T.~Enevoldsen, E.~Fa{\ss}hauer, T.~Fleig, O.~Fossgaard, A.~S.~P.~Gomes,
  E.~D.~Hedeg{\aa}rd, T.~Helgaker, J.~Henriksson, M.~Ilia{\v{s}}, Ch.~R.~Jacob,
  S.~Knecht, S.~Komorovsk{\'y}, O.~Kullie, J.~K.~L{\ae}rdahl, C.~V.~Larsen,
  Y.~S.~Lee, H.~S.~Nataraj, M.~K.~Nayak, P.~Norman, G.~Olejniczak, J.~Olsen,
  J.~M.~H.~Olsen, Y.~C.~Park, J.~K.~Pedersen, M.~Pernpointner, R.~di~Remigio,
  K.~Ruud, P.~Sa{\l}ek, B.~Schimmelpfennig, A.~Shee, J.~Sikkema,
  A.~J.~Thorvaldsen, J.~Thyssen, J.~van~Stralen, S.~Villaume, O.~Visser,
  T.~Winther, and S.~Yamamoto (available at
  \url{https://doi.org/10.5281/zenodo.2253986}, see also
  \url{http://www.diracprogram.org})}\BibitemShut {NoStop}%
\bibitem [{\citenamefont {Yanai}\ \emph {et~al.}(2004)\citenamefont {Yanai},
  \citenamefont {Tew},\ and\ \citenamefont {Handy}}]{yanai2004new}%
  \BibitemOpen
  \bibfield  {author} {\bibinfo {author} {\bibfnamefont {T.}~\bibnamefont
  {Yanai}}, \bibinfo {author} {\bibfnamefont {D.~P.}\ \bibnamefont {Tew}},\
  and\ \bibinfo {author} {\bibfnamefont {N.~C.}\ \bibnamefont {Handy}},\
  }\bibfield  {title} {\bibinfo {title} {A new hybrid exchange--correlation
  functional using the {Coulomb-attenuating method (CAM-B3LYP)}},\ }\href@noop
  {} {\bibfield  {journal} {\bibinfo  {journal} {Chem. Phys. Lett.}\ }\textbf
  {\bibinfo {volume} {393}},\ \bibinfo {pages} {51} (\bibinfo {year}
  {2004})}\BibitemShut {NoStop}%
\bibitem [{\citenamefont {Dyall}(2016)}]{dyallHv2z}%
  \BibitemOpen
  \bibfield  {author} {\bibinfo {author} {\bibfnamefont {K.~G.}\ \bibnamefont
  {Dyall}},\ }\bibfield  {title} {\bibinfo {title} {Relativistic double-zeta,
  triple-zeta, and quadruple-zeta basis sets for the light elements{ H--Ar}},\
  }\href@noop {} {\bibfield  {journal} {\bibinfo  {journal} {Theor. Chem.
  Acc.}\ }\textbf {\bibinfo {volume} {135}},\ \bibinfo {pages} {128} (\bibinfo
  {year} {2016})}\BibitemShut {NoStop}%
\bibitem [{\citenamefont {Dyall}(2004)}]{dyallRuv3z}%
  \BibitemOpen
  \bibfield  {author} {\bibinfo {author} {\bibfnamefont {K.~G.}\ \bibnamefont
  {Dyall}},\ }\bibfield  {title} {\bibinfo {title} {Relativistic double-zeta,
  triple-zeta, and quadruple-zeta basis sets for the 5$d$ elements {Hf--Hg}},\
  }\href@noop {} {\bibfield  {journal} {\bibinfo  {journal} {Theor. Chem.
  Acc.}\ }\textbf {\bibinfo {volume} {112}},\ \bibinfo {pages} {403} (\bibinfo
  {year} {2004})},\ \bibinfo {note} {revision K.G. Dyall and A.S.P. Gomes,
  Theor. Chem. Acc. (2009) 125:97.}\BibitemShut {Stop}%
\bibitem [{\citenamefont {Ilia{\v{s}}}\ and\ \citenamefont
  {Saue}(2007)}]{iliavs2007infinite}%
  \BibitemOpen
  \bibfield  {author} {\bibinfo {author} {\bibfnamefont {M.}~\bibnamefont
  {Ilia{\v{s}}}}\ and\ \bibinfo {author} {\bibfnamefont {T.}~\bibnamefont
  {Saue}},\ }\bibfield  {title} {\bibinfo {title} {An infinite-order
  two-component relativistic {Hamiltonian} by a simple one-step
  transformation},\ }\href@noop {} {\bibfield  {journal} {\bibinfo  {journal}
  {J. Chem. Phys.}\ }\textbf {\bibinfo {volume} {126}},\ \bibinfo {pages}
  {064102} (\bibinfo {year} {2007})}\BibitemShut {NoStop}%
\bibitem [{\citenamefont {Schimmelpfennig}(1996)}]{schimmelpfennig1996amfi}%
  \BibitemOpen
  \bibfield  {author} {\bibinfo {author} {\bibfnamefont {B.}~\bibnamefont
  {Schimmelpfennig}},\ }\bibfield  {title} {\bibinfo {title} {{AMFI}, an atomic
  mean-field spin-orbit integral program},\ }\href@noop {} {\bibfield
  {journal} {\bibinfo  {journal} {Stockholm University}\ } (\bibinfo {year}
  {1996})}\BibitemShut {NoStop}%
\end{thebibliography}%

\end{document}